\documentclass[12pt,A4paper]{article}

\usepackage{setspace}
\usepackage{amsthm}

\usepackage{amsmath}
\usepackage{multirow}
\usepackage{booktabs}
\usepackage{tikz}
\usetikzlibrary{matrix,positioning}
\usetikzlibrary{shapes.geometric, arrows, positioning}
\usepackage{pgfplots}
\pgfplotsset{compat=1.16}
\usepgfplotslibrary{fillbetween}
\pgfplotsset{ticks=none}
\usetikzlibrary{calc}
\usepackage[round]{natbib}
\usepackage{geometry}
\usepackage{rotating}
\usepackage{tabulary}
\usepackage{lmodern}
\usepackage[T1]{fontenc}
\usepackage{hhline} 
\usepackage{graphicx}
\usepackage[hidelinks]{hyperref}
\usepackage{xcolor}
\usepackage{xurl}
\usepackage{float}
\usepackage{mdframed}
\usepackage{tikz}
\mdfsetup{%
  backgroundcolor=gray!10,
  roundcorner=5pt,
  framemethod=tikz
}
\usepackage{enumitem}
\setlist[itemize]{noitemsep, topsep=0pt}
\usepackage{url}

\hypersetup{
    colorlinks,
    linkcolor={red!50!black},
    citecolor={blue!70!black},
    urlcolor={blue!70!black}
}
\usepackage{threeparttable} 
\usepackage{cleveref}
\newcommand{\citetposs}[1]{\citeauthor{#1}'s~(\citeyear{#1})}

\title{\fontsize{20pt}{22pt}\selectfont From Model Design to Organizational Design:\\ Complexity Redistribution and Trade-Offs in Generative AI\footnote{Author names are in alphabetical order. Please direct correspondence to ssamila@iese.edu.}}
\author{Sharique Hasan \\ \textit{\fontsize{12pt}{14pt}\selectfont Duke University} \and Alexander Oettl \\ \textit{\fontsize{12pt}{14pt}\selectfont Georgia Institute of Technology} \and Sampsa Samila \\ \textit{\fontsize{12pt}{14pt}\selectfont IESE Business School}}
\date{June 10, 2025}

\begin{document}

\maketitle

\begin{abstract}
This paper introduces the Generality-Accuracy-Simplicity (GAS) framework to analyze how large language models (LLMs) are reshaping organizations and competitive strategy. We argue that viewing AI as a simple reduction in input costs overlooks two critical dynamics: (a) the inherent trade-offs among generality, accuracy, and simplicity, and (b) the redistribution of complexity across stakeholders. While LLMs appear to defy the traditional trade-off by offering high generality and accuracy through simple interfaces, this user-facing simplicity masks a significant shift of complexity to infrastructure, compliance, and specialized personnel. The GAS trade-off, therefore, does not disappear but is relocated from the user to the organization, creating new managerial challenges, particularly around accuracy in high-stakes applications. We contend that competitive advantage no longer stems from mere AI adoption, but from mastering this redistributed complexity through the design of abstraction layers, workflow alignment, and complementary expertise. This study advances AI strategy by clarifying how scalable cognition relocates complexity and redefines the conditions for technology integration.
\end{abstract}



\newpage

\newpage
\doublespacing

\newpage
\section{Introduction}\label{introduction}

Approximately forty seconds after a client interaction, a Morgan Stanley wealth-management advisor accesses a browser interface displaying three outputs: a structured meeting summary, a draft follow-up email, and updated CRM records. Tasks that previously required thirty minutes are now completed in seconds via a tool called \textit{Debrief}, saving millions of minutes annually and reallocating effort toward client engagement \citep{son_morgan_2024, openai_ms_evals_2024}. This apparent simplicity masks a complex infrastructure of regulatory data pipelines, model-risk frameworks, and vector-indexed repositories governed by a multidisciplinary oversight group. The contrast between surface usability and hidden complexity illustrates a broader tension this paper investigates \citep{openai_ms_evals_2024, chan_morgan_2024}.

A similar configuration appears at Klarna. In early 2024, its customer-service bot handled 2.3 million chats in one month---the workload of 700 agents---with a projected \$40 million profit increase \citep{klarna2024}. In marketing, generative tools reduced campaign timelines from six weeks to seven days and cut agency costs by 37\% while improving content quality and personalization \citep{Graham2024KlarnaAI}. As with Morgan Stanley, this performance depends on back-end systems, real-time decision engines, experimentation platforms, and compliance guardrails, that remain invisible to end-users \citep{tsruya_klarna_realtime_2023}. 

Current conceptualizations frame AI progress primarily as a reduction in the cost of specific activities or inputs, such as prediction or expertise. However, this narrow view overlooks two critical dynamics evident in these examples. First, AI models must navigate inherent tradeoffs among multiple competing dimensions rather than optimizing a single metric. Second, the deployment of AI redistributes complexity across different stakeholders within and beyond the organization, fundamentally reshaping firm boundaries and determining strategic success.

A long-standing principle in model-building---across statistical, physical, and algorithmic domains---holds that designers must manage tradeoffs among three goals: \emph{Generality}, \emph{Accuracy}, and \emph{Simplicity} (GAS). Specifically:

\begin{itemize}
\item \textbf{Generality} refers to a model's ability to perform across diverse contexts.
\item \textbf{Accuracy} denotes the degree to which outputs align with empirical observations.
\item \textbf{Simplicity} concerns both the effort required for users to understand, apply, or interact with the model and the underlying complexity of the model itself.
\end{itemize}

In principle, improving generality and accuracy requires increased model complexity, which, in turn, reduces simplicity. Historically, AI systems optimized for one or two dimensions by sacrificing the third. The empirical puzzle is why generative AI models such as ChatGPT appear to violate this constraint: they exhibit high generality and accuracy while remaining simple to use.

This management of complexity occurs through technical and organizational abstraction. While the user interacts with a uniform and accessible interface, underlying complexity is shifted to other layers, engineers, infrastructure teams, and regulatory systems who manage and encapsulate this complexity. This redistribution of complexity is central to understanding how organizations scale advanced AI technologies without violating the GAS tradeoff.

\paragraph{Abstraction layers hide complexity and reshape organizational boundaries.}
The public release of OpenAI’s ChatGPT on November 30, 2022, marked a significant inflection point. Its rapid and widespread initial adoption \citep{Hu2023} leading now to 43\% US workers using similar tools by April 2025 \citep{Hartley2025} can be attributed to a distinctive combination: broad generality and reasonable accuracy, delivered through the simplicity of a natural language interface. This interface functions as a powerful abstraction layer. Shielding users from underlying technical complexity enables an experience that appears to optimize generality, accuracy, and simplicity simultaneously, even though these qualities typically involve tradeoffs in model design.

The abstraction layer changes how people interact with the system. Users and developers do not need detailed knowledge of each other's work. Instead, they rely on a shared interface that standardizes inputs and outputs. This separation is much stronger than in earlier machine learning tools and breaks the direct connection between those building the model and those applying it. As a result, new boundaries form within and across organizations. Although the system feels simple to use, the underlying complexity still exists. It is shifted to other parts of the organization, where it is handled by specialists in areas like prompt engineering, data governance, and AI ethics.\footnote{A similar dynamic occurs in other technical domains. For example, software engineers working in JavaScript do not need to understand assembly language, and those writing in assembly do not require knowledge of semiconductor physics. Abstraction layers modularize expertise and support scalability.}

\paragraph{Large language models reduce the unit cost of generalist cognition.}
Just as steam power reduced the cost of mechanical work and semiconductors lowered the cost of arithmetic operations, contemporary large language models (LLMs) such as OpenAI’s GPT-4.5 and Google’s Gemini 2.5 reduce the cost of general-purpose cognitive tasks. When the cost of a general-purpose input declines, established economic dynamics follow: (i) usage expands rapidly, (ii) previously uneconomical applications become feasible (as in Klarna’s personalized marketing campaigns), (iii) substitute activities (such as routine writing or customer-service staffing) decrease in value, and (iv) complementary capabilities (including relationship-building, data quality, specialized tooling, and oversight capacity) increase in strategic importance.

\vspace{0.9cm}
\noindent The cases of Morgan Stanley and Klarna represent early indicators of broader structural change. This paper argues that understanding this redistribution of complexity is the central strategic challenge for organizations in the AI era. While users enjoy unprecedented simplicity, the burden of complexity shifts to technical infrastructure, compliance frameworks, and specialized new roles within the firm and to new markets outside of the firm. The GAS tradeoff re-emerges, not at the user's screen, but in the form of an ``accuracy ceiling'', operational bottlenecks, and new organizational risks.

Consequently, competitive advantage will not be determined by simply adopting AI to cut costs. It will be forged by mastering the architecture of this redistributed complexity: by designing intelligent workflows, building complementary human expertise, and making deliberate strategic choices about where to operate on the Generality-Accuracy frontier. This paper provides a conceptual framework for leaders to navigate these tradeoffs and rethink the architecture of work and strategy in an age of seemingly free, generalist intelligence. The goal is not to provide a prescriptive implementation manual but to offer a conceptual framework for rethinking the architecture of work, organizational design, and competitive strategy in the age of increasingly inexpensive generalist intelligence.


\section{The Generality-Accuracy-Simplicity Tradeoff}
\label{the-generality-accuracy-simplicity-tradeoff}

\subsection{The GAS Trade-off in Model Design}

The introduction outlined how the potential of generative AI stems from its ability to navigate a persistent tension in model design: the trade-off among generality, accuracy, and simplicity.

Throughout history, humans have developed models to understand and manage the complexity of natural and social systems. These models have ranged from intuitive analogies, such as the planetary model of atomic structure, to advanced computational simulations for forecasting climate dynamics or epidemic spread. Contemporary artificial intelligence, including large language models (LLMs), represents the latest stage in this trajectory. While these systems differ in scale and architecture, they share a common set of features. In particular, the need to simplify and structure complex realities in ways that support reasoning and action.

A foundational constraint in this modeling tradition is that no model can perfectly replicate the phenomena it represents. Instead, effective models must balance competing priorities across three dimensions: generality, accuracy, and simplicity, referred to here as the GAS framework. Generality (G) describes the breadth of contexts, domains, or tasks a model can address. Accuracy (A) reflects how outputs align with observable reality. Simplicity (S) refers not only to the cognitive and technical ease with which users can interact with the model but also to the degree to which underlying complexity is abstracted or redistributed to other system layers. In practice, achieving high generality and accuracy often requires substantial internal complexity. Simplicity depends less on eliminating complexity and more on managing where it resides. Because organizations cannot maximize these dimensions simultaneously, model design involves deliberate trade-offs among them.

\subsubsection{Historical Foundations of the GAS Framework}

The trade-off among generality, accuracy, and simplicity has been articulated in several influential discussions of modeling across the sciences. The formulation presented here aligns with Thorngate’s Postulate in Social Psychology \citep{Thorngate1976}, which asserts that no theory can simultaneously be general, accurate, and simple. Specifically, any attempt to optimize one or two of these dimensions entails compromises along the third. \citet{Weick2015} elaborated on this insight by noting that increasing a theory’s precision or scope often diminishes its comprehensibility or parsimony. This perspective mirrors the central tension captured by the GAS framework, in which gains in accuracy may reduce either generality or simplicity. In both formulations, efforts to represent the complexity of real-world phenomena risk obscuring key insights or limiting practical applicability. Conversely, simplifying a model to enhance clarity or usability can obscure substantial domain-specific variation.

In the field of ecology, \citet{Levins1966} introduced a triad of model properties: \emph{generality}, \emph{realism}, and \emph{precision}. Levins defined realism as the extent of mechanistic detail embedded in a model. Adapting this framework to the social sciences, we follow Thorngate in renaming this dimension as \emph{simplicity}, which we define as reflecting both the cognitive effort required to understand and use the model and the structural complexity of the model itself. This broader interpretation highlights an inverse relationship between mechanistic detail and simplicity: as realism increases, simplicity decreases due to heightened interpretive demands and greater internal complexity.\footnote{Realism corresponds to high mechanistic detail; simplicity corresponds to low mechanistic detail. They represent opposite ends of the same axis, framed with different emphases.} For both Levins and Thorngate, tractability is the limiting factor. Models that attempt to maximize both generality and accuracy tend to become overly complex, thereby reducing their usefulness and manageability and forcing trade-offs across the three dimensions.

More broadly, in the philosophy of science, \citet{Kuhn1977}, in a lecture delivered in 1973, identified five characteristics that guide the selection of a preferred scientific theory: accuracy, scope (or generality), simplicity, consistency (both internal and with other accepted theories), and fruitfulness (in terms of generating new lines of inquiry). The first three correspond directly to the dimensions articulated in Thorngate’s Postulate and the GAS framework. Kuhn also acknowledged the inherent trade-offs among these criteria. As he observed, ``Individually the criteria are imprecise... In addition, when deployed together, they repeatedly prove to conflict with one another; accuracy may, for example, dictate the choice of one theory, scope the choice of its competitor'' \citep[][p. 322]{Kuhn1977}. This observation reinforces the general insight that efforts to improve one dimension often come at the expense of others, a dynamic central to scientific theory-building and model design in applied domains such as artificial intelligence.

\subsubsection{Empirical Illustration: Domain-Specific vs. General-Purpose Models}

\begin{figure}[H]
    \centering
    \includegraphics[width=\linewidth]{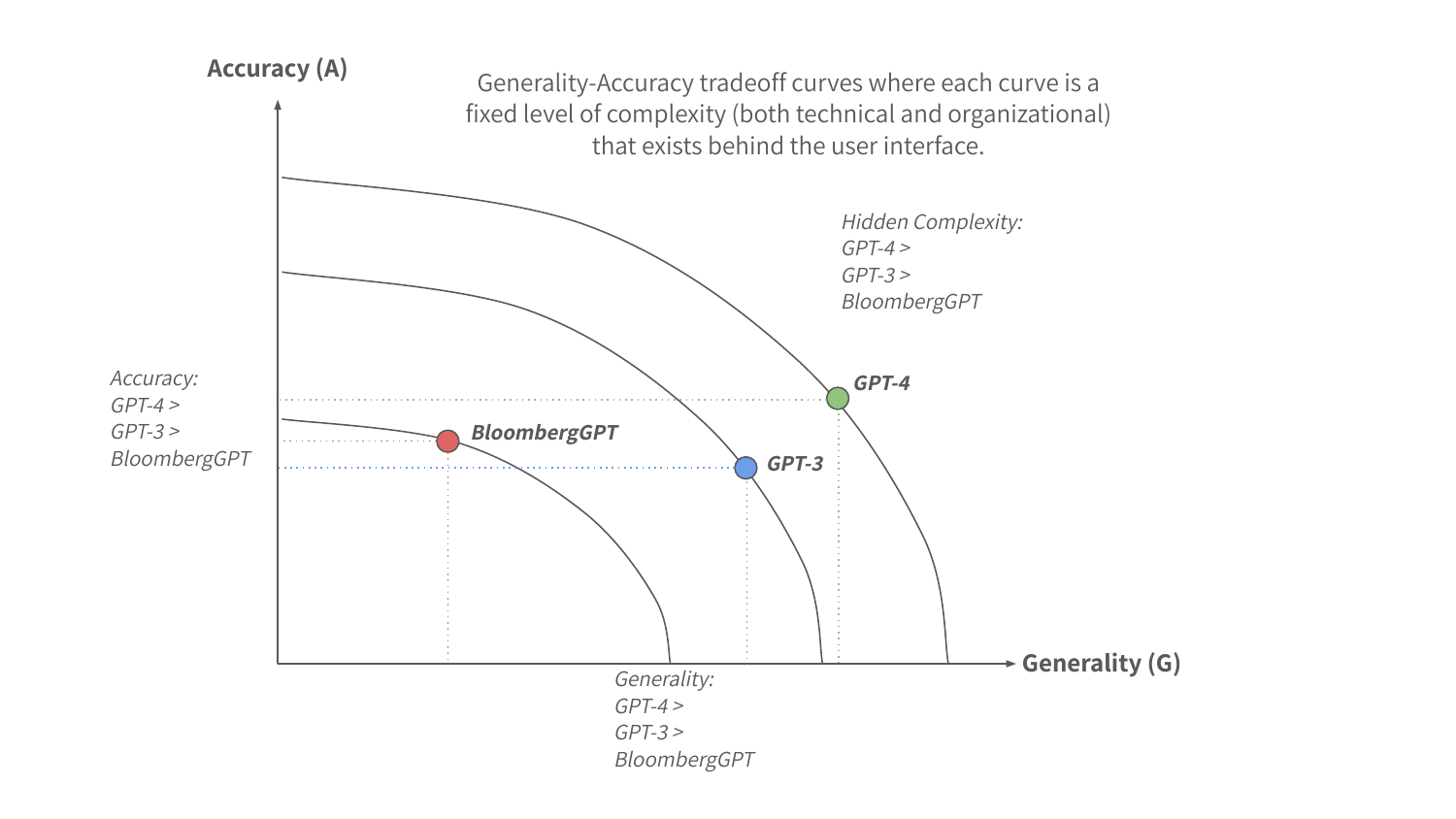}
    \caption{Generality–Accuracy trade-off curves at fixed levels of hidden complexity. Each curve represents a constant technical and organizational complexity that is not visible to the end-user. While specialized models such as BloombergGPT prioritize accuracy within narrow domains, more complex general-purpose models (e.g., GPT-4) can exceed them in both accuracy and generality by operating at higher levels of hidden complexity.}
    \label{fig:GAtradeoff}
\end{figure}

An illustrative example of the GAS trade-off in the context of AI language models is BloombergGPT, a 50-billion-parameter model trained primarily on Bloomberg’s proprietary corpus of financial news, regulatory filings, and pricing data. By restricting the training domain to finance, the model’s designers accepted a reduction in generality (G) to improve accuracy (A) on domain-specific tasks \citep{BloombergGPT2023, Wu2023}. Compared to more general-purpose models such as OpenAI’s GPT-3, which served as the foundation for early versions of ChatGPT, BloombergGPT achieved higher accuracy on financial applications, including equity sentiment analysis, bond question answering, and other finance-specific benchmarks. This increase in task-specific accuracy was accompanied by a corresponding decrease in generality, with the model performing less effectively outside its specialized domain. We depict this trade-off in Figure \ref{fig:GAtradeoff}.\footnote{The accuracy–generality frontiers are the level-sets of the GAS trade-off function  $S = 1 - (A^\alpha + G^\alpha)^\beta$,  where $\alpha>0$ governs the steepness of the penalty (higher $\alpha$ makes the drop-off near extreme $A$ or $G$ sharper) and $\beta>0$ controls the overall convexity of the curve (larger $\beta$ bows it outwards, reflecting stronger diminishing returns). Solving for $A$ at a fixed simplicity level $S$ gives 
$A = \bigl((1 - S)^{1/\beta} - G^\alpha\bigr)^{1/\alpha}$.  By letting $G$ vary over its feasible range and plotting the corresponding $A$, one can trace a single frontier for a given complexity budget; successive outer curves correspond to less simple $S$ models (i.e., more hidden technical and organizational complexity).}

OpenAI’s next model, GPT-4, provides a contrasting case that challenges the typical constraints of the GAS trade-off. By incorporating significantly greater complexity through expanded model parameters, increased data diversity, and higher training compute, GPT-4 improved accuracy (A) and generality (G) at the same time. It outperformed domain-specific models such as BloombergGPT on specialized financial tasks while maintaining strong performance across a wide range of general-purpose benchmarks \citep{Li2023}. A substantial increase in system-level complexity enabled these gains. However, interface design and abstraction largely preserved user-facing simplicity (S), allowing non-expert users to interact with the model without engaging with its internal technical structure. In this context, complexity refers to the total effort required to create, understand, deploy, and maintain the model. It encompasses the underlying technical architecture, the operational demands of infrastructure and governance, and the specialized cognitive work required by developers and system designers, with each component creating distinct economic niches for new jobs and services.

\section{Redistributing Complexity Through Abstraction}

\subsection{Classical Machine Learning and the Limits of Generality}

\begin{figure}[H]
    \centering
    \includegraphics[width=\linewidth]{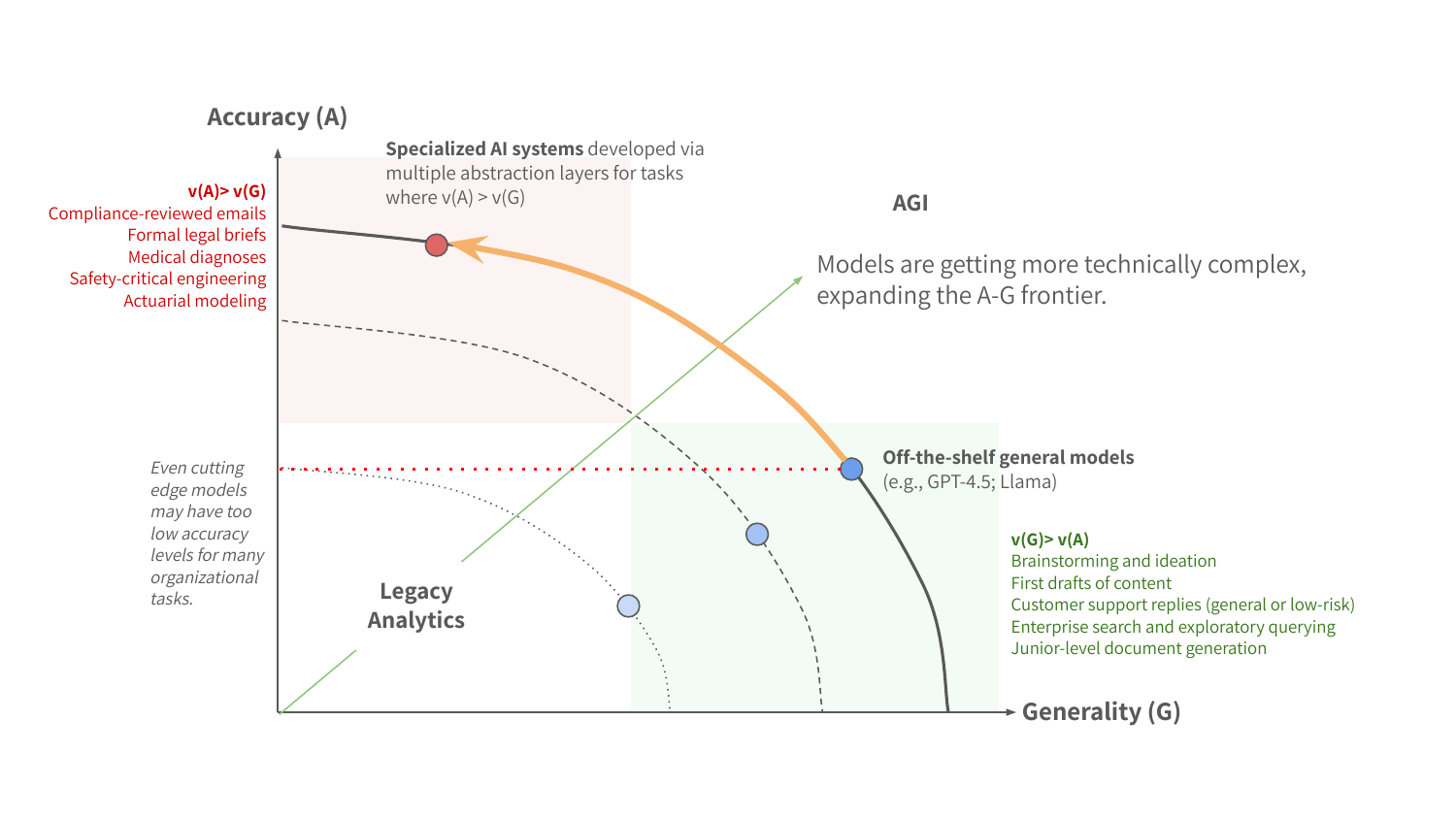}
    \caption{Tasks vary in their accuracy and generality demands, shaping where they fall along the A–G frontier. High-stakes, domain-specific work favors specialized systems, while broad, low-risk tasks can be handled by general-purpose models.}
    \label{fig:GtoA}
\end{figure}

Classical predictive machine learning models occupy a distinct region of the GAS trade-off space. Models such as gradient-boosted trees for churn prediction, XGBoost classifiers for credit risk, or logistic regressions for fraud detection are designed to optimize performance on narrowly defined tasks. They exhibit low generality (G), as their applicability is confined to specific problem domains. For example, a credit risk model is not expected to summarize earnings calls or engage with unrelated content. Within their targeted scope, however, these models often achieve high accuracy (A), frequently outperforming human judgment or rule-based systems. At the same time, their interfaces are typically simple to use, with outputs that are easily interpretable by business users or operational teams. Traditional supervised learning, therefore, represents a canonical case of high accuracy and low generality, with simplicity preserved at the user level.

\subsection{Abstraction as a Mechanism for Complexity Management}
\label{separation-of-complexity}

As introduced earlier, a primary strategy for managing the GAS trade-off and enabling the appearance of high generality and accuracy through a simple user interface is abstraction. This approach is a direct application of a long-standing principle in interaction design known as Tesler's Law, or the Law of Conservation of Complexity \citep{TeslerInterview2006, Yablonski2024}. Formulated in the 1980s by computer scientist Larry Tesler, the law states that for any system, there is an inherent amount of complexity that cannot be eliminated, only redistributed. The critical question is who must deal with it: the user, or the designers and developers. Tesler argued that for mass-market software, it is economically advantageous to make the software internally more complex if it saves millions of users even a minute of effort each day. In this context, abstraction refers to the deliberate concealment of underlying technical and operational complexity from the end-user.


The chat interface of models such as ChatGPT illustrates the dynamic. It conceals a neural architecture of extraordinary complexity, consisting of hundreds of billions of parameters, behind a simple and intuitive natural language prompt. This represents the ``surface ease'' layered over the ``hidden machinery'' observed in the Morgan Stanley and Klarna cases. In this context, abstraction layers function as contracts. Individuals on either side of the interface interact based on shared expectations and standardized inputs and outputs without requiring detailed knowledge of the system's internal workings.

Before the emergence of highly generalized large language models (LLMs), the development of machine learning systems often required users to possess technical expertise in machine learning methods and a detailed understanding of the target domain. This dual requirement arose partly because the abstraction layer between the model and its application context was relatively thin. Effective deployment typically depended on extensive domain-specific feature engineering, custom data preprocessing, and iterative adaptation. As a result, machine learning development teams frequently needed to work near business units to co-develop and refine practical solutions.

However, these abstractions that separate one part of the system from another are never flawless. As influential software engineer Joel Spolsky articulated in his `Law of Leaky Abstractions,' all non-trivial abstractions are, to some degree, leaky \citep{Spolsky2002Leaky}. The underlying complexity an abstraction is meant to hide inevitably `leaks through,' creating performance issues, unexpected failures, and errors that can only be debugged by understanding the lower-level system the abstraction was meant to conceal.

\subsubsection{Abstraction in Software Development as an Analogy}

A longitudinal analysis of programming paradigm evolution illustrates how complexity has been progressively redistributed through abstraction. In the earliest phases of computing, developers interacted directly with hardware using machine language. This approach required deep familiarity with specific hardware architectures and often resulted in systems that were difficult to maintain and prone to error. The introduction of assembly language marked the first significant abstraction layer. Replacing binary instructions with human-readable mnemonics improved code interpretability, although programmers still had to manage hardware-specific details.

As computational technologies matured, higher-level programming languages were introduced to abstract away increasingly large portions of low-level complexity. These abstractions enabled developers to concentrate more fully on problem-domain conceptualization rather than on implementation mechanics. Functional programming languages, for example, simplify code reasoning by representing computation in terms of mathematical functions. Object-oriented programming advanced abstraction by structuring programs around interacting ``objects'' that encapsulate internal complexity and expose only well-defined interfaces. Each successive shift in paradigm redistributed complexity away from the user and toward the underlying system, allowing broader accessibility and more scalable development.
This progression of abstraction finds a clear parallel and a significant inflection point in the development of machine learning. In classical machine learning, effective modeling often requires a deep integration of statistical expertise with domain-specific knowledge. Feature engineering, a critical and labor-intensive phase, necessitated that developers understand the operational logic of the domain they were modeling. For example, to train a churn-prediction gradient-boosting tree or a fraud-detection XGBoost model, teams had to analyze customer behavior patterns or fraud mechanisms in detail and then manually construct features and thresholds that captured those patterns. This close coupling of modeling skill and domain insight often required continuous collaboration between technical teams and business stakeholders. As a result, classical machine learning systems typically occupied the high-accuracy, low-generality region of the GAS triangle, with performance gains realized through deep specialization.

Contemporary large language models (LLMs) represent a continuation of the abstraction trajectory, with a notable shift in where complexity resides. These systems allow technical and non-technical users to articulate computational goals in natural language, relying on the model to interpret and operationalize those instructions. This reduces the cognitive and procedural burden traditionally associated with programming, abstracting away hardware and syntax and the need for tightly coupled domain expertise. Yet, while LLMs simplify the user experience, they introduce new forms of complexity at other system levels. Challenges related to interpretability, backend reliability, and alignment with human values persist, illustrating that the GAS trade-off remains operative even as its terms shift.

\subsubsection{Redistributing Complexity Across Stakeholders}

From the perspective of the \textbf{end-user}, models such as GPT-4 may appear to defy the GAS constraint. Generality and accuracy both improve, while the user interface remains unchanged. This apparent resolution does not eliminate the simplicity dimension. Instead, simplicity is preserved at the surface by shifting complexity to deeper system layers. The interface feels intuitive because abstraction has relocated the cognitive and technical burdens elsewhere in the organizational and technical stack.

From the perspective of \textbf{developers}, the GAS balance reasserts itself. Increases in generality and accuracy are achieved at the expense of simplicity, reflected in rising complexity across model training, evaluation, serving infrastructure, and governance. This hidden complexity can be formally modeled; research in complex systems engineering quantifies this `structural complexity' based on the number of components, their interfaces, and their topological arrangement. Critically, this research provides empirical validation that development effort does not scale linearly with this complexity. Instead, it grows super-linearly, meaning the operational and financial costs absorbed by the organization accelerate as the back-end systems become more intricate \citep{Sinha2013, Sinha2016}. 

This hidden complexity is a direct manifestation of what \citet{sculley2015hidden} termed ``hidden technical debt'' in machine learning systems. They famously illustrated that only a tiny fraction of the code in many real-world ML systems is devoted to the core learning or prediction algorithms. The vast majority is the surrounding infrastructure required for configuration, data collection, data verification, feature extraction, and process management. This debt accrues in tangible forms, including massive amounts of ``glue code'' written to get data in and out of general-purpose packages, sprawling and brittle ``pipeline jungles'' for data preparation, and extensive, error-prone system configurations.


This redistribution of complexity from the user to the organization is not merely theoretical; it is visible in labor market shifts where firms adopting AI significantly increase their hiring of managers to oversee these new, complex systems \citep{Alekseeva2024}.

The scale of this hidden complexity is not trivial. Bank of America’s virtual assistant, Erica, which presents a simple conversational interface to 42 million clients, has required over 50,000 performance updates by its data science team since its launch. This continuous, behind-the-scenes effort is what preserves the `surface ease' for the user while the system's capabilities in Generality and Accuracy expand \citep{bofa2024erica}.

This redistribution of complexity is a deliberate strategic choice. For instance, when off-the-shelf AI coding tools proved inadequate for modernizing its legacy software, Morgan Stanley developed its own tool, DevGen.AI. While commercial tools offer high Generality, the bank opted to absorb the complexity of building an in-house solution to achieve the specific Accuracy required for its unique, and sometimes proprietary, coding languages. The tool saved developers 280,000 hours in its first few months, demonstrating a clear return on the investment in managing this complexity internally \citep{bousquette2025morgan}.

Further emphasizing the strategic nature, this principle of managing complexity through abstraction is now central to corporate strategy in diverse industries. Yum! Brands, parent to KFC and Pizza Hut, is consolidating its global restaurant technologies into a single proprietary SaaS platform called Byte by Yum!. This platform provides franchisees with a simplified and integrated suite of AI-driven tools for ordering, point of sale, kitchen optimization, and inventory management. By creating this abstraction layer, Yum! absorbs the immense backend complexity to deliver operational simplicity and `advantaged economics' to its thousands of franchise partners \citep{yum2025byte}.

\subsubsection{Development of Large-Scale Complexity}

The development of models of immense complexity has been enabled by a central insight articulated by \citet{Sutton2019} in his essay ``The Bitter Lesson'' and empirically supported by the scaling laws demonstrated in \citet{Kaplan2020}. Sutton argues that the primary driver of progress in artificial intelligence is not human ingenuity or understanding but the systematic exploitation of computational resources. Rather than crafting highly specialized, interpretable systems, developers have achieved superior performance by building larger models and training them on more data for extended periods. The \citet{Kaplan2020} study demonstrated that model performance improves predictably as a function of three variables: dataset size, model size, and training duration. These improvements occur independently of developers' detailed understanding of the resulting system. Computational scale, not interpretive clarity, governs model effectiveness. This approach has enabled simultaneous gains in generality and accuracy, even though the internal workings of the models often remain opaque.\footnote{The AI research company Anthropic has actively investigated the internal mechanisms of large language models, publishing a series of interpretability studies. These efforts remain ongoing. See: https://www.anthropic.com/research}

The performance gains enabled by large-scale computation allow developers of foundational large language models to focus on building and refining general-purpose architectures rather than mastering the specifics of every downstream application. In contrast to traditional machine learning workflows, which often require close alignment between technical modeling and domain expertise, LLM developers concentrate on the model's intrinsic complexity: its architecture, training process, and broad capabilities. The assumption is that sufficient scale and generality will allow the model to perform effectively across diverse tasks, including those not explicitly anticipated during development. This shift reflects a reorientation from task-specific optimization to the cultivation of general cognitive capacity through scale.

\begin{figure}[H]
    \centering
    \includegraphics[width=\linewidth]{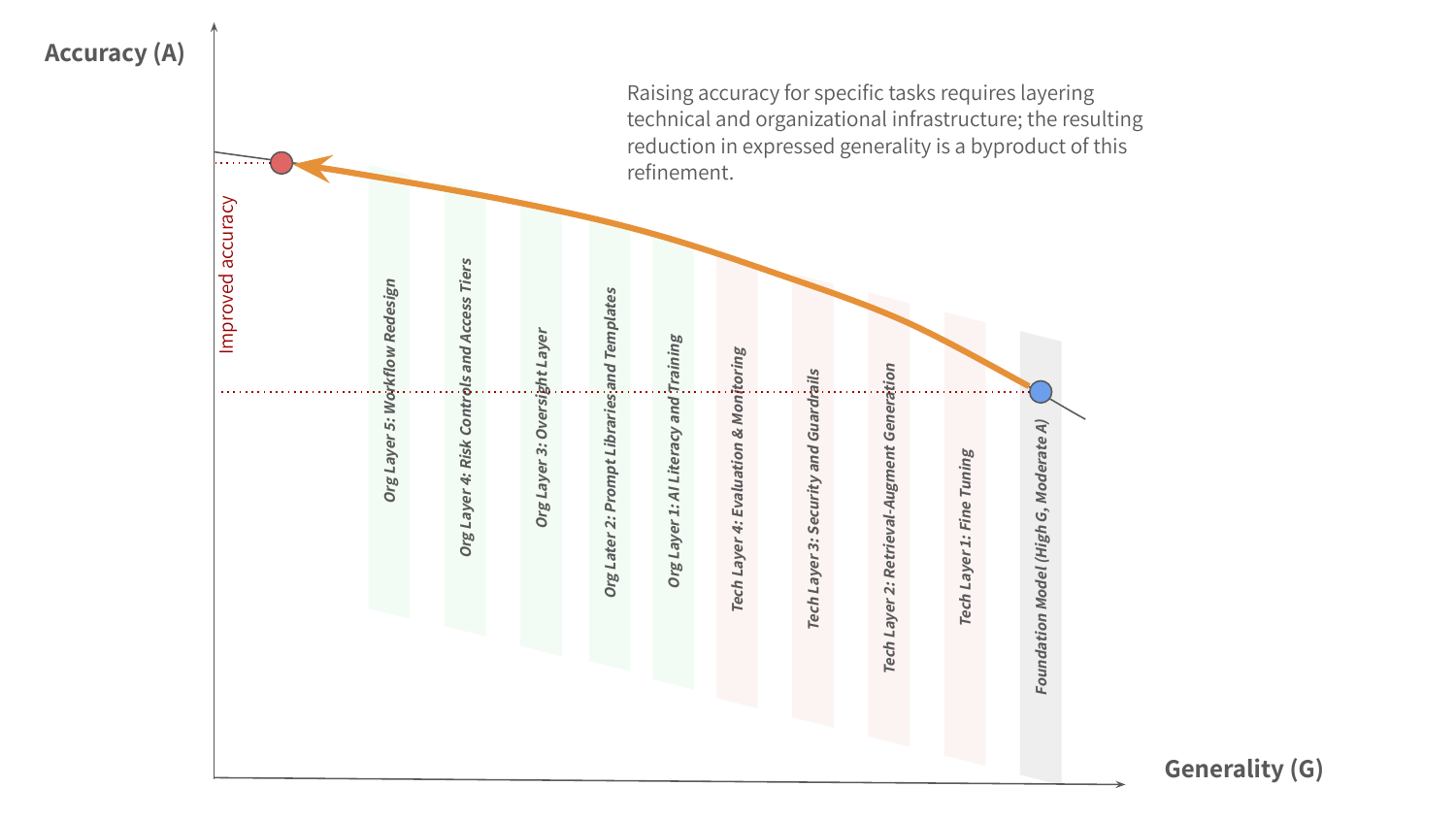}
    \caption{To achieve high task-specific accuracy, organizations layer technical and organizational infrastructure between the foundation model and the user. This structured refinement increases accuracy while narrowing the model’s expressed generality.}
    \label{fig:ValueChain}
\end{figure}

However, it is important to recognize that this powerful abstraction layer reconfigures rather than eliminates the need for expertise. Organizational boundaries shift as critical domain knowledge relocates to the ``last-mile'' application layers. Deep understanding of specific business problems becomes essential for tasks such as prompt engineering, the design and deployment of Retrieval Augmented Generation (RAG) systems,\footnote{Retrieval Augmented Generation (RAG) is analogous to providing a language model with an open-book exam. Rather than relying solely on pre-trained knowledge, the system retrieves relevant and up-to-date information from specified sources, such as internal documents or curated databases, and uses that information to generate more accurate and contextually grounded responses. See \citet{Lewis2020}.} the fine-tuning of models for specialized domains,\footnote{Fine-tuning refers to adapting a general-purpose model to a specific task or dataset. While it improves performance within a narrowly defined scope, it often reduces generality by making the model more sensitive to the structure and assumptions of the fine-tuning data.} and the evaluation of model outputs to ensure alignment with organizational accuracy and compliance requirements. Although the interface appears simple, practical implementation still depends on substantial human expertise in the downstream layers of the system.

In short, scaling does not eliminate the GAS constraint. Users benefit from a seemingly costless experience, but developers absorb the burden through hidden complexity rather than cognitive load. This cost extends beyond direct financial and computational requirements. It includes the ongoing challenge of managing large-scale models' subtle and often unpredictable behaviors. This unpredictability is a symptom of a core challenge in modern machine learning known as underspecification: a single training pipeline can produce many different models that perform identically on standard metrics but behave very differently in real-world scenarios due to arbitrary choices like the random seed used during training \citep{damour2020}.


In some cases, models express high confidence in incorrect responses \citep{VarianceLabs2025}. Addressing this issue requires the development of external heuristics, model-specific calibration techniques, and additional product safeguards. These efforts increase operational complexity behind the scenes, even as end-users interact with an interface that remains simple and intuitive.

A critical constraint in designing and deploying AI systems is the directional asymmetry between specialization and generalization. It is comparatively straightforward to adapt generalist models for specific domains through fine-tuning methods such as Low-Rank Adaptation (LoRA; \citet{Hu2021}) or retrieval augmentation strategies such as RAG \citep{Lewis2020}. However, extending a domain-specific or specialist model to general-purpose use is considerably more difficult. This asymmetry reflects both empirical observations and theoretical challenges in continual learning. In particular, efforts to expand a specialist model by introducing data from unrelated domains often result in catastrophic forgetting, where previously acquired capabilities are lost \citep{Luo2025, Parisi2019}. Consequently, generalist models offer a more flexible foundation. They allow organizations to develop customized applications without initiating model training from scratch, whereas repurposing narrow models typically requires reconstructing the system from a more general architecture \citep{MalaquiasJunior2025}.

From the developer's perspective, the simplicity axis in the GAS framework is often the first to collapse. Each additional layer of user-facing functionality, such as extended context windows, multimodal capabilities, or improved safety responses, depends on substantial engineering investment. These advances require managing large-scale infrastructure, including multi-GPU training clusters, sharded model checkpoints, retrieval augmentation systems, red-team evaluations, and compliance reviews increasingly resembling regulatory filings. What appears to the end-user as a seamless interaction, e.g., typing a question and receiving an answer, is, in reality, the product of a complex system consisting of orchestration scripts, data lineage trackers, and automated regression testing pipelines. The apparent simplicity of the chat interface is enabled by a corresponding increase in hidden complexity at the system level.

\citetposs{Sutton2019} \emph{Bitter Lesson} explains why this burden continues to grow: historical evidence suggests that raw computational power consistently outperforms human ingenuity, making it more effective to scale models using additional data and compute rather than relying on hand-engineered features. \citetposs{Kaplan2020} scaling laws provide empirical support, showing that prediction error decreases in a predictable power-law relationship as model parameters, dataset size, and training compute increase. However, these same scaling trends also impose growing costs. Each marginal improvement in accuracy or generality requires disproportionately greater investments in infrastructure, energy consumption, and safety mechanisms. The apparent simplicity of the user interface, often reduced to a single input line, is sustained only because simplicity for developers has been traded for scale-driven complexity.

This separation of concerns, enabled by the abstraction that large language models provide, extends beyond internal team structure and has broader implications for industry structure. It supports a market-level division where some firms concentrate on the capital-intensive development and maintenance of foundational models, focusing on advancing generality and model capability. Other firms operate at the application layer, building domain-specific products and services on top of these general-purpose platforms without incurring the full cost or complexity of foundational model development.

However, the value extracted by these application-layer firms is not guaranteed by the capabilities of the foundational model alone. Empirical research shows that upgrading from GPT-3.5 to GPT-4 yields only modest performance gains unless users are explicitly trained to collaborate with AI systems. In many cases, training interventions have a greater impact on task performance than the underlying model upgrade itself \citep{Li2024}. This finding underscores the importance of user capability and AI literacy in determining whether the apparent simplicity of the interface translates into realized generality and accuracy in downstream applications. As a result, the broader AI ecosystem is evolving toward a layered structure in which foundational model providers and application builders occupy increasingly specialized and complementary roles.

\section{Constraints on Accuracy in Generalist Models}

\subsection{LLMs as Pattern Matching: The Inherent Accuracy Ceiling}
\label{llms-as-statistical-pattern-matching-systems-capabilities-and-constraints}

The previous section examined how abstraction layers enable large language models (LLMs) to deliver a user experience characterized by high generality and simplicity, even as system complexity increases behind the scenes. However, this managed complexity, combined with the foundational architecture of LLMs, introduces inherent constraints, particularly along the accuracy dimension of the GAS framework. This limitation can be described as an ``accuracy ceiling'': a practical boundary on the precision these systems can achieve, especially as their scope and generality continue to expand.

The fundamental architecture of LLMs as pattern-matchers means they can generate plausible but incorrect inferences, a phenomenon observed by experts in highly technical fields. This issue is exacerbated by the problem of underspecification: the problem is not just pattern-matching, but that there are many possible patterns a model could learn, all of which look equally good during training. While these models are treated as equivalent, they may rely on different spurious correlations or ``shortcuts'', leading them to fail in  different and unpredictable ways on the same real-world task. The choice between a model that generalizes correctly and one that relies on a spurious feature can thus be determined by something as arbitrary as the random seed used for initialization \citep{damour2020}.

This technical limitation is not merely academic, but a recurring challenge for practitioners. As one scientist noted when probing the models on familiar topics: ``I realize that, even with the deep research features and citations, it's making a bunch of incorrect inferences that likely arise from certain concepts (words, really) co-occurring in documents but are actually physically not causally linked or otherwise fundamentally connected.''\footnote{https://news.ycombinator.com/item?id=44162499}

A large-scale 2024 study evaluating nine different state-of-the-art LLMs using formal verification methods found that at least 62\% of the AI-generated C programs contained security vulnerabilities \citep{tihanyi2024secure}. A similar study found that top models frequently overlook security issues, resulting in the creation of vulnerable code \citep{wang2024secure}. The flaws are not trivial; the most common vulnerabilities across all models were severe memory-related errors like NULL pointer dereferences and buffer overflows, which can lead to system crashes and security exploits. 

A 2025 study evaluating the creativity of 14 major LLMs found that while the models often surpassed average human performance, their output rarely reached elite levels \citep{haase2025creativity}. Furthermore, the study revealed significant intra-model variability, where identical prompts could produce outputs ranging from subpar to exceptional. This demonstrates that even for general cognitive tasks, the accuracy of LLM outputs is neither consistently high nor reliable, reinforcing the need for organizational systems to manage this variance.

Crucially, the strategic importance of the accuracy ceiling is not absolute, but relative to the performance of the incumbent human-led process. Human experts face their own accuracy ceiling, limited by cognitive load and situational complexity. In a landmark study, a frontier LLM demonstrated `superhuman' diagnostic ability, consistently outperforming expert attending physicians with real patient data from an emergency room \citep{brodeur2024superhuman}. The model's advantage was most pronounced during initial triage, the moment of highest uncertainty, where it proved significantly more accurate than its human expert counterparts. From a GAS perspective, the model's immense hidden technical complexity allows it to better manage the situational complexity that limits human performance. This re-frames the organizational challenge: the goal becomes managing a more accurate, but still imperfect, AI-driven process, shifting the human's role from creator to critical overseer.

\subsection{The Accuracy Ceiling in Generalist AI Systems}

The statistical architecture that enables large language models (LLMs) to achieve high levels of \textbf{generality} and interface-level \textbf{simplicity} also imposes constraints on \textbf{accuracy}, the third dimension of the GAS framework. LLMs function primarily as advanced statistical pattern recognition systems. As \citet{Blender2021} describe, they are effectively ``stochastic parrots,'' which are systems that excel at identifying, synthesizing, and reproducing complex linguistic and conceptual patterns, often surpassing human intuition in pattern recognition. LLMs internalize statistical relationships within the training distribution and produce outputs that exhibit contextual coherence and apparent relevance by training on massive corpora of text, code, and structured data. 

This feature of LLMs as pattern matchers is demonstrated by a recent study which finds that even top-tier LLMs suffer performance losses exceeding 60\% on elementary reasoning problems when a subtle condition is changed. It was enough to change the word ``traveling'' to ``floating'' for describing motion on water. This is consistent with the models reciting familiar solution templates instead of reasoning from the new premises \citep{yan2025recitation}.

The accuracy ceiling is no longer merely theoretical; it is now being measured in practical domains. An example is a large-scale experiment by \citet{Hackenburg2025Scaling} that tested persuasive messages from 24 language models of varying sizes on over 25,000 people. The study found ``sharply diminishing returns'' in persuasiveness (a specific form of accuracy) as model size (a proxy for model complexity) increased. Frontier models (at the time of the experiment) like GPT-4-Turbo and Claude-3-Opus were only slightly more persuasive than models an order of magnitude smaller, with this advantage almost entirely explained by their superior ability to perform basic ``task completion'' (i.e., generating coherent, on-topic messages). This empirically validates a core tenet of the GAS framework: after a baseline of functional accuracy is met, further increases in complexity for generalist models yield diminishing returns. Because current frontier models have already mastered basic task completion, the results suggest that strategic advantage shifts from scaling model size to optimizing workflows and alignment.

A large-scale 2025 study simulating over 200,000 conversations found that the performance of even state-of-the-art models like GPT-4.1 and Gemini 2.5 Pro degrades by an average of 39\% when shifting from a single, fully-specified instruction to a multi-turn, underspecified conversation \citep{laban2025llms}. The researchers found that models `get lost' when they make premature assumptions and then fail to recover from early mistakes. This academic finding gives rigorous validation to a common occurrence, where an early AI error cascades, rendering a conversation unsalvageable and forcing the user to start over with a fresh context. This forces a redistribution of complexity directly onto the user, who must learn strategies like manually summarizing and cleaning context to manage the model's accuracy limitations.

Despite their performance, large language models (LLMs) lack semantic understanding and explanatory depth. The limitation is evident in tasks requiring formal logic or rigorous deductive reasoning. For instance, although LLMs may successfully solve some mathematical problems, they often struggle to produce formal proofs or to maintain logical consistency across extended chains of reasoning \citep{Mirzadeh2024, Petrov2025}. The emergent nature of large models further compounds this issue. Their scale and training complexity introduce capabilities and failure modes that remain only partially understood, often resulting in unpredictable or difficult-to-evaluate behaviors \citep{Bommasani2022}.

This limitation in symbolic and procedural logic is illustrated by experiments where models were explicitly provided with the correct algorithm to solve a puzzle. Apple researchers \citep{Shojaee2025illusion} found that all state-of-the-art Large Reasoning Models (LRMs) face a ``complete accuracy collapse'' to zero beyond a certain, measurable complexity threshold. Even with the step-by-step solution logic in the prompt, the models' performance did not improve, and they failed at roughly the same complexity point. This inability to reliably execute prescribed steps highlights that the core deficit lies not just in finding a solution, but in the fundamental mechanics of consistent logical execution, reinforcing their nature as pattern-matchers rather than true reasoners \citep{Shojaee2025illusion}. This provides empirical support for the existence of an accuracy ceiling inherent in current architectures.

These failure modes are not incidental; they represent concrete manifestations of the pressure exerted on the accuracy dimension when generality and simplicity are maximized. This inherent limitation gives rise to what can be described as the generality–accuracy (G–A) frontier, where expanding a model’s scope often comes at the cost of reduced precision unless accompanied by a substantial increase in complexity. One of the clearest expressions of this dynamic is the phenomenon of hallucinations, e.g., plausible-sounding but factually incorrect outputs. This challenge is central to the practical implications of the accuracy ceiling.

Recent empirical findings further illustrate this trade-off. OpenAI’s technical report on its latest reasoning models documents that as model generality increased, factual accuracy unexpectedly declined \citep{OpenAI2025}. Specifically, the OpenAI o3 model hallucinated answers to 33 percent of questions in the internal PersonQA benchmark, roughly double the 16 percent hallucination rate recorded by its less general predecessor, OpenAI o1. The report also notes that OpenAI o3 produced more total claims than OpenAI o1, which resulted in a higher absolute number of accurate statements and a greater number of hallucinated responses. Similarly, the smaller OpenAI o4-mini model hallucinated in 48 percent of PersonQA instances and 79 percent of SimpleQA items \citep{OpenAI2025}. These results reinforce the G–A trade-off: increases in general reasoning capability do not inherently reduce error rates. In some cases, they may increase them unless significant alignment and control mechanisms are applied.

The location of this G-A frontier in the live economy is now becoming empirically visible. Recent data from millions of user interactions shows that AI adoption peaks in high-skill but not the highest-stake professions, such as software development. Usage drops off sharply for the highest-wage, highest-liability occupations like physicians and for occupations requiring extensive advanced degrees, where the cost of an error is unacceptably high. This provides suggestive real-world evidence that organizational deployment strategies are implicitly governed by the accuracy ceiling, delegating tasks to AI only where the cost per mistake remains within acceptable boundaries \citep{Handa2025}.

\subsubsection{Alignment Trade-offs and the Limits of Fine-Tuning}

While intended to improve specific aspects of \textbf{Accuracy (A)} and safety, the alignment process often introduces what has been described as an ``alignment tax.'' \citet{Ouyang2022} introduced this concept in the context of aligning GPT-3 with human feedback. While the alignment procedure improved user-friendliness---a desirable dimension of \textbf{A}---it also reduced performance on several established benchmarks. This trade-off illustrates that alignment efforts can enhance targeted aspects of accuracy while diminishing overall \textbf{Generality (G)} or other components of \textbf{A}.

Subsequent studies have further quantified these effects. Reinforcement-based alignment methods, such as Reinforcement Learning from Human Feedback (RLHF), have been shown to impair knowledge retention and reduce creative output, both key attributes of \textbf{G} \citep{Perez2022, Mohammadi2024}. \citet{Perez2022} found that increased RLHF training, although aimed at improving specific forms of \textbf{A}, can degrade model performance on unrelated tasks. This phenomenon, known as inverse scaling, underscores alignment interventions' non-linear and sometimes counterproductive effects.

\citet{Mohammadi2024}, in their study Creativity Has Left the Chat, similarly, report that intensive content filtering and value alignment reduce output diversity. Models begin to converge toward repetitive or overly cautious responses, referred to as ``attractor states.'' In this case, the general creative capability is constrained to increase safety or reliability, highlighting a trade-off between \textbf{G} and narrowly scoped \textbf{A}.

Finally, alignment can distort a model’s internal calibration. Post-training adjustments to enhance specific types of accuracy may lead to overconfident outputs, even when the model is incorrect. This reduces the reliability of its probability estimates and obscures uncertainty signals that engineers use for monitoring and risk mitigation \citep{VarianceLabs2025}. These effects illustrate that alignment does not eliminate the GAS trade-off; it reconfigures it, often shifting complexity to less visible aspects of model behavior.

In a study on fine-tuning LLMs for specialized business tasks, \citet{zhang2023balancing} found that using purely in-domain data improved task-specific accuracy but caused a `notable decline' in the model's general abilities. Their solution was to develop a complex data-blending strategy to consciously manage the Generality-Accuracy balance. This operational-level challenge perfectly illustrates the GAS framework in practice: achieving useful accuracy requires significant investment in data strategy and custom evaluation, forms of hidden complexity, while actively trading off a degree of generality.

\subsubsection{Infrastructure Strategies for Improving Accuracy}

Engineers seeking to improve \textbf{Accuracy (A)} in large language models can typically do so by sacrificing one of the other two dimensions in the GAS framework. Domain-specific fine-tuning increases accuracy for narrowly defined tasks but reduces \textbf{Generality (G)}. Alternatively, developers may implement complex retrieval or verification mechanisms, constraining generality and significantly increasing system-level complexity. A prominent example of the latter approach is Retrieval Augmented Generation (RAG), which enhances factual accuracy by grounding model outputs in external documents \citep{Lewis2020}. This strategy introduces substantial backend complexity, even as the user interface remains unchanged.

The organizational implications of such complexity are evident in applied settings. Kaitlin Elliott, Head of Firm-wide Generative AI Solutions at Morgan Stanley, described this challenge in a collaboration with OpenAI: ``Based on all the questions we input and outputs we’re getting, we’d sit with OpenAI and say, `What can we change about our retrieval methods to help the accuracy we need at Morgan Stanley?' '' \citep{openai_ms_evals_2024}. Her account illustrates how gains in accuracy require detailed, iterative coordination around infrastructure and workflow design, much of which remains invisible to end-users.

However, the use of grounding mechanisms inherently constrains an LLM’s expressed \textbf{Generality (G)} within a given interaction by limiting its outputs to the scope of the retrieved context. These systems also introduce additional architectural and operational complexity, which can reduce user-facing \textbf{Simplicity (S)}. Moreover, they remain vulnerable to hallucinations and factual inaccuracies when the retrieval process is flawed or when the retrieved content is incomplete or misleading \citep{Huang2024}. 

A 2025 study evaluating leading AI models, including OpenAI's GPT-4o and Anthropic's Claude 3.5 Sonnet, tested their ability to answer 185 patent law questions while being grounded in a single, specific casebook. This ``closed-book'' approach, which is the foundational principle of RAG, was expected to significantly reduce errors. However, the performance was poor. The study's authors graded a substantial portion of responses as ``unacceptable'', containing substantial errors of law, for GPT-4o (26\%), Claude 3.5 (14\%), and Google's NotebookLM (31\%). The performance of Google's NotebookLM is particularly revealing, as it is a tool specifically designed for this kind of task. Despite this specialization, its 31\% ``unacceptable" rate was the highest of the three \citep{LarrimoreOuellette2025}.

This accuracy challenge is magnified when organizations, having specialized a model with internal documents to increase task-specific Accuracy at the expense of Generality, then attempt to regain that Generality by granting the agent access to the open web. Recent research on this ``safety devolution'' shows that such a move can lead to a significant loss of safety accuracy, with agents becoming more biased and more likely to comply with harmful requests \citep{yu2025safety}. The effect was so pronounced that retrieval-augmented agents built on ``safe'', aligned LLMs often produced more harmful content and exhibited more bias than their uncensored, non-retrieval counterparts. 

This limitation underscores a key insight within the GAS framework: the simultaneous achievement of high generality and simplicity often requires compromises in \textbf{Accuracy (A)} or explanatory precision, especially in domains where formal reasoning and verifiable correctness are critical. 

As a result, human critical judgment, domain-specific expertise, and ongoing oversight play a central role in the responsible deployment of LLMs. These systems achieve their full potential only when integrated within structured workflows and management protocols that account for their technical constraints and epistemic limitations.

For organizational leaders, the core lesson is a strategic one: any deployment must select a position along the Generality–Accuracy (G–A) frontier. Expanding \textbf{Generality (G)} increases the risk of hallucinations and potential reputational harm. In contrast, enhancing \textbf{Accuracy (A)} through alignment procedures or retrieval-based techniques such as Retrieval Augmented Generation (RAG) often introduces additional cost and complexity while also narrowing the effective generality of the model’s outputs in a given context. As a result, forward-looking organizations distribute tasks across differentiated workflows. Broad, general-purpose models are reserved for internal tasks such as ideation and drafting. In contrast, externally facing or compliance-sensitive functions are routed through narrower, highly aligned systems, sometimes supported by curated retrieval mechanisms.

The central question is no longer which model performs best in isolation but how much expressed \textbf{G} a given workflow can support when optimizing for \textbf{A} in a specific use case. Strategic deployment, therefore, depends on model selection, workflow architecture, and risk tolerance.

\section{Operational Limits: Latency, Cost, and Risk}

\subsection{Operational Constraints: Latency and Cost}
\label{sec:latency_cost}

While the statistical architecture of large language models (LLMs) imposes an inherent ``accuracy ceiling,'' as discussed above, the practical deployment of these systems is subject to additional constraints that arise from the same complexity that underpins their broad \textbf{Generality (G)} and aspirational \textbf{Accuracy (A)}. Beyond the conceptual trade-offs defined by the GAS framework, two key operational limitations are particularly salient: latency, referring to the time required to generate a response, and operating cost, which captures the computational expense associated with each interaction. Both factors are direct consequences of increased model complexity. As models grow in size and capability, their demands on computational infrastructure intensify, resulting in slower response times and higher per-query costs. These constraints limit where and how organizations can deploy such models at scale.

\paragraph{Latency as a hidden accuracy limit.}
Deploying larger models often results in significantly increased response times. For example, GPT-4 has a higher response time than GPT-3.5 , introducing a delay of two seconds for a typical chat response \citep{latif2024p}. These delays have measurable effects on user engagement. Rasa, a leading conversational AI platform, observed that higher latency from larger models led to increased user dissatisfaction and premature termination of conversations. By transitioning to smaller models, Rasa achieved faster response times, improving user experience in real-time chat applications \citep{Rasa2025}. Open-source benchmarks reinforce this finding. Engineers at Databricks report that LLaMA-2 with 70 billion parameters responds approximately two times more slowly than its 13 billion–-parameter counterpart, even after optimization through parallel processing \citep{Agarwal2023Databricks}.

These observations suggest that latency imposes a practical upper bound on the accuracy that can be effectively utilized. If responses arrive too slowly, the user-facing \textbf{Simplicity (S)} that makes chat interfaces attractive begins to erode, undermining the overall utility of high-performing but slow models.

\paragraph{Cost as the deployment governor.} Inference costs increase with the number of model parameters and the volume of tokens processed. A recent analysis by Epoch AI shows that the cost of achieving GPT-4–level performance on PhD-level science benchmarks has declined sharply, by approximately \textbf{40-fold per year}, with some tasks exhibiting cost reductions as high as 900-fold. However, this varies considerably across benchmarks \citep{epoch2025llminferencepricetrends}. Despite these trends, analysts at Andreessen Horowitz estimate that, under current cloud pricing, generating one million tokens using a frontier model costs approximately 20 to 40 times more than doing so with a mid-tier alternative \citep{Appenzeller2024LLMflation}.

Thus, even as absolute costs decline, relative differences remain substantial. Smaller, less complex models consistently offer lower latency and operational costs than their frontier-scale counterparts. These persistent cost differentials reinforce the need for a careful deployment strategy, particularly in contexts requiring high-frequency, low-margin inference.

\paragraph{Managerial takeaway.} Latency and cost jointly constitute a second practical frontier that limits how far any organization can advance along the dimensions of \textbf{Generality (G)} and \textbf{Accuracy (A)}. Each workflow should, therefore, be evaluated using a simple decision grid: \emph{How quickly must the system respond?} and \emph{What is the acceptable cost per response?} In many cases, the optimal solution is a hybrid architecture, where lightweight models address routine queries and more complex models are reserved for high-stakes or ambiguous tasks. Realizing the benefits of large-scale hidden complexity thus requires both technical optimization and sound economic judgment.

These operational constraints---latency, cost, and risk---impose practical boundaries on every workflow, forcing organizations to make strategic trade-offs along the Generality–Accuracy frontier. In the aggregate, such choices redefine the accuracy ceiling and delimit the extent to which tasks can be automated with current model architectures.

\section{Organizational and Strategic Implications}
\label{impact-on-work-the-accuracy-ceiling-in-practice}

\subsection{Organizational Implications of the GAS Trade-Off}

The preceding sections have established a foundational premise: all models, including those underpinning contemporary generative AI systems, are constrained by the Generality, Accuracy and Simplicity (GAS) trade-off. We have shown that abstraction layers, such as the natural language interfaces of large language models (LLMs), enable users to interact with these systems through seemingly simple, general-purpose workflows. This design conceals substantial technical and operational complexity, allowing for the simultaneous appearance of high generality and simplicity.

However, this complexity has been redistributed, not eliminated. It usually shifts to engineering teams, infrastructure, and organizational processes. This shift creates its own compounding costs, as the maintenance of ML systems over time is notoriously difficult and expensive. As a result, the trade-off reasserts itself in less visible but no less significant ways \citep{sculley2015hidden}. Most notably, it emerges as an ``accuracy ceiling,'' which constrains how reliably these models can perform, especially in domains requiring precise reasoning or factual rigor.

This friction is acutely visible in the legal services industry. As of mid-2025, corporate clients report not seeing expected cost savings from their law firms’ adoption of generative AI. While clients expect AI’s Generality to curtail bills, firms struggle to demonstrate this, citing their own investment costs which offset efficiency gains. This impasse highlights a core GAS tension: the firm absorbs the hidden complexity, which then conflicts with the client’s expectation of cost-based Simplicity \citep{Muscavage2025clarity}.

\subsubsection{Expanded Generality Through Interface-Level Simplicity}

The emergence of large language models represents the most significant expansion of generality in the history of digital tools. A single conversational interface can now perform tasks as varied as translating policy documents, generating TypeScript code, summarizing a corporate wiki, or drafting a professional apology, all without requiring the user to acquire new syntactic knowledge. The capabilities of an email agent illustrate this shift: through a prompt-based interface, Google’s Gemini system now performs functions such as message labeling, triage, and response generation, functions that previously relied on multiple specialized applications. Similar patterns are also present in software engineering. Models can generate tens of thousands of lines of code across diverse technology stacks within short time frames.

The wide generality of Large Language Models (LLMs) is supported by both predictive and empirical research. One study predicted that around 80\% of the U.S. workforce could see at least 10\% of their work tasks affected by LLMs \citep{Eloundou2023}. More recent empirical work confirms this breadth, finding that 36\% of occupations already use LLMs for at least a quarter of their associated tasks \citep{Handa2025}. For instance, a field experiment at Procter \& Gamble found that individuals using generative AI were able to produce new product solutions of the same quality as two-person cross-functional teams. The study also found that AI helped both technical and commercial professionals produce more balanced solutions \citep{DellAcqua2025Cybernetic}. This is not confined to specific sectors; adoption is varied, with industries from `Information Services' to `Real Estate' and `Construction' all reporting usage rates of over 40\% \citep{Hartley2025}.

From the perspective of the GAS framework, this transformation reflects a profound redistribution of complexity. The generality is embedded within the model weights, enabling users to access expanded capabilities through a static, simplified interface. The result is a striking increase in domain breadth, achieved without requiring the user to confront the underlying system complexity. This pattern of diffusion, which has been documented across nearly all industries since the early 2010s, creates a new set of organizational imperatives focused not on developing the technology itself, but on managing its deployment \citep{Alekseeva2021}.

\subsubsection{Accuracy Constraints and Metacognitive Burden}

Yet this expanded reach remains constrained by a persistent shortfall in \textbf{Accuracy (A)}. Hallucinations, defined as confident but unfounded assertions, are the most well-known manifestation. Additional limitations include inconsistent responses to identical prompts, fragility in performance when the context window is altered, and forms of reward optimization that substitute superficial solutions for substantive ones. These issues all stem from a foundational characteristic of large language models: they are statistical pattern completion systems, not reasoning engines. 

As a result, effective use requires users to sustain calibrated trust in their domain knowledge and to evaluate outputs critically rather than rely on the model's surface fluency. The most common reported use of generative AI is to `get the task done quicker', with only 16\% of users stating the tool `completed the task' for them \citep{Hartley2025}. This reveals a widespread human-in-the-loop workflow, where AI serves as an assistant to be supervised rather than an autonomous agent. This introduces a significant metacognitive burden, as users must continually regulate their confidence and monitor for plausible but incorrect outputs \citep{Tankelevitch2024}.

The problem is not merely that a given model might be wrong, but that the training process itself is likely to be underspecified, meaning that identically trained models can learn different spurious shortcuts and exhibit entirely different failure modes. Therefore, a user is not just learning to collaborate with a single AI, but with an unstable process that can produce predictors with varying and unpredictable biases \citep{damour2020}.

This burden is tangible, often manifesting as a shift in the nature of work from pure creation to verification and debugging. As practitioners on technical forums note, using AI can become less productive when one is ``spending all your time debugging its outputs''. This has led to the observation that for many developers, ``LLM-coding has largely just shifted `time spent typing' to `time spent reviewing' ''.\footnote{https://news.ycombinator.com/item?id=44162373} 

One challenge is managing the ongoing conversational context. As one practitioner noted, once the context is `poisoned' by an error, ``it will not recover, you need to start fresh with a new chat''.\footnote{https://news.ycombinator.com/item?id=43991384} This constant need for vigilance, context management, and knowing when to abandon a failing interaction is a primary component of the metacognitive burden placed on the user. 

The core value proposition of AI tools, therefore, depends on this trade-off being favorable; the effort of verification and vigilance must be significantly less than the effort of creation from scratch. To highlight this concern, the authors of a 2025 legal education study noted that the time they spent as domain experts carefully reviewing the AI's lengthy and often subtly flawed responses was  greater than the time it would have taken to simply write the correct answers themselves \citep{LarrimoreOuellette2025}.

This shift is evident in detailed collaboration logs. In an experiment involving ad creation, humans paired with AI agents focused 23\% more on messaging about content generation and 20\% less on direct text editing, as the AI handled much of the iterative writing \citep{Ju2025Collaborating}. This reflects a reallocation of human effort from routine execution to higher-level guidance and collaboration with the AI system.

Indeed, some influential practitioners now frame this as a new core competency. When an LLM produces mediocre or even low-quality code, the ability for a senior developer to correct it effectively is seen as a skills issue to be managed, rather than a fundamental flaw of the technology itself \citep{Ptacek2025Nuts}.

The rising importance of this human-led oversight is reflected in how the skill requirements for managers are evolving. Recent evidence shows that as firms integrate AI, they seek managers with stronger cognitive and interpersonal skills, such as collaboration, creativity, and data analysis, while the need for routine administrative skills like scheduling and budgeting declines. The manager's role is potentially shifting from a doer of automatable tasks to a curator and director of AI-augmented workflows \citep{Alekseeva2024}.

\subsubsection{Risk, Liability, and the Emergence of AI Error Insurance}

The tangible impacts of this accuracy deficit are visible across a range of professional environments. For instance, an AI-powered support bot for a software development platform recently prompted significant user backlash after providing incorrect information about company policies. This failure resulted from the model fabricating details that did not exist, a direct consequence of hallucination. More concerning are emerging reports indicating that, despite improvements in specific capabilities, newer AI systems do not necessarily exhibit reduced rates of factual error. In some cases, advanced models designed for reasoning tasks have demonstrated higher hallucination rates than their predecessors on specific benchmarks \citep{metz2025hallucinations, OpenAI2025}.

A recent randomized controlled trial involving upper-level law students engaged in realistic legal tasks illustrates this dynamic. The study found that an advanced AI reasoning model (OpenAI's o1-preview) improved the analytical depth of responses but also produced a greater number of hallucinated legal citations compared to both a Retrieval Augmented Generation (RAG)-based AI tool and human work unaided by AI. The RAG-based system substantially reduced hallucinations, suggesting that specific architectural strategies, such as retrieval augmentation, can mitigate accuracy challenges even when the underlying model remains prone to error \citep{Schwarcz2025}. These findings highlight that the ``accuracy ceiling'' is not a fixed threshold but a dynamic and ongoing constraint in LLM deployment. They also reinforce a central implication of the GAS framework: beyond a certain point, improvements in generality can be achieved only by accepting trade-offs in reliability.

The financial and reputational risks associated with accuracy limitations in AI systems have grown sufficiently acute to spur the development of new insurance products. In May 2025, Lloyd's of London insurers introduced specialized coverage for losses arising from malfunctioning AI tools. The policies, developed in partnership with Y Combinator-backed startup Armilla, cover legal fees and damages resulting from court claims against companies whose AI systems underperform relative to expectations \citep{Harris2025}. Several recent incidents have underscored the relevance of such protections: Air Canada was compelled to honor erroneous discounts offered by its chatbot; Virgin Money faced backlash after its AI tool flagged the word ``virgin'' as inappropriate; and courier company DPD temporarily shut down its AI-powered customer service after it issued profane messages and criticized the company. This development illustrates how firms are beginning to formalize and financially quantify the risks of the persistent accuracy ceiling inherent in current AI technologies.

In parallel, the Big Four accountancy firms are building AI-assurance audits that adapt traditional financial-statement techniques to verify the safety, bias, and regulatory compliance of AI systems. Deloitte calls such assurance `critical to AI adoption', and PwC plans to launch a dedicated practice `soon' \citep{Kissin2025bigfourft}. These moves shift part of the accuracy burden from builders to third-party professionals, reinforcing the GAS insight that complexity is re-distributed rather than eliminated.

\subsubsection{Practitioner Frameworks for Managing the GAS Trade-Off}

The challenge of managing the accuracy ceiling is the central problem facing AI product builders. In response, practitioners have developed operational strategies that serve as applied solutions to the GAS trade-off. These strategies implicitly acknowledge that competitive advantage comes not from waiting for perfect models, but from building systems that explicitly manage their inherent variance.

This complexity extends even to the initial assessment of a tool's capabilities. For instance, a 2024 GitHub study claimed its Copilot assistant produced code that was significantly more functional, readable, reliable, maintainable, and concise \citep{Bauer2024}. However, these claims were challenged by independent software developers who argued the study used an overly simplistic coding task, one likely to be represented in the training data, and relied on a definition of ``code errors'' that excluded functional bugs in favor of stylistic issues \citep{Cimpianu2024, Claburn2024}.

This dispute exemplifies a form of redistributed complexity: organizations must not only verify the AI's output but also develop the in-house expertise to scrutinize vendor methodologies and metrics. Trusting a vendor's ``accuracy'' claims without this critical oversight becomes a strategic risk. In response, practitioners have developed operational strategies that serve as applied solutions to the GAS trade-off.

\citet{Grobman2025}, for instance, outlines four distinct approaches for builders that directly map onto different methods of managing redistributed complexity and selecting a position on the Generality-Accuracy frontier. The strategies fall into two broad categories:

\begin{itemize}
\item Autonomous Systems: These approaches aim to deliver a product with high user-facing Accuracy (A) and Simplicity (S) by investing heavily in hidden technical and organizational complexity. This can involve pursuing near-perfect determinism in a narrow domain (sacrificing Generality (G) for maximum A) or ensuring a system is accurate enough for low-stakes tasks where a level of errors is tolerable.
\item Human-in-the-Loop Systems: These approaches accept the model's Accuracy (A) limitations and explicitly redistribute the complexity of verification to human roles. This can involve end-user verification, where copilots assist knowledge workers who perform the final accuracy check , or provider-level verification, where a service provider absorbs the complexity of review to deliver a polished, reliable final product to the client.
\end{itemize}

These practitioner models confirm the central thesis of the GAS framework: unreliability is a fundamental constraint, and value is created by strategically managing, rather than eliminating, the trade-offs between Generality, Accuracy, and Simplicity.

\subsection{Mapping Workflows to the GAS Frontier}

\citetposs{Dhar2016} Decision-Automation Map makes this trade-off explicit. By plotting \emph{predictability} against the \emph{cost per mistake}, the framework defines an upward-sloping frontier that delineates tasks suitable for automation from those requiring human oversight. If one reinterprets ``predictability'' as a proxy for accuracy, the diagram can be seen as a concrete cross-section of the GAS space. Tasks such as spam filtering or display-ad bidding, which carry relatively low accuracy requirements and low consequences for errors, have long been delegated to automated systems. In contrast, high-stakes applications such as autonomous vehicle control occupy regions of high error cost, where even modest residual inaccuracies render full automation impractical.

The strategic allocation of AI resources is visible at the highest levels of enterprise. JPMorgan Chase, with its \$18 billion technology budget, exemplifies this mapping of workflows to the G-A frontier. For low-risk, high-volume tasks, it uses AI and automation to reduce call center servicing costs and anticipates a 10\% headcount reduction in operational roles. For mid-risk, high-skill work, it deploys 'copilot' tools like Connect Coach to augment its 7,600 wealth advisors, improving their productivity by sourcing discussion materials 90\% faster. In high-stakes domains, it has reduced the cost of client verification in its investment bank by 40\% using highly specialized AI applications \citep{abrego2025jpmorgan}.

Mapping contemporary knowledge work onto this surface helps explain the uneven practical impact of large language models. LLMs produce substantial productivity gains and broaden access to expertise in domains where the cost of error is low, such as drafting marketing copy, generating initial code scaffolds, or supporting early-stage brainstorming. Junior professionals may experience two- to three-fold increases in output while senior colleagues shift into supervisory roles. LLMs tend to create an overproduction dynamic in mid-risk contexts, including customer support replies, compliance-reviewed emails, or enterprise search. Multiple plausible outputs are generated rapidly, requiring human evaluators to sift, select, and refine. Knowledge workers transition from creators to curators in these cases, spending time adjudicating machine-generated proposals. LLMs function primarily as adjuncts in high-risk applications, such as formal legal briefs, medical diagnoses, or safety-critical avionics systems. Their use is limited to information retrieval or exploratory ideation, as stringent accuracy requirements and liability concerns prevent their full integration into core decision-making processes.

\subsection{Empirical Evidence of Productivity Effects}
\label{empirical-evidence-of-productivity-effects}

The productivity gains observed in field experiments reflect the GAS framework's prediction that increases in generality and user-facing simplicity facilitate rapid adoption, while limitations in accuracy lead to uneven impacts across tasks and skill levels. Since large language models optimize for Generality and Simplicity, measurable productivity improvements tend to appear first in domains where the consequences of accuracy errors are minimal.

The initial real-world deployments of GPT-powered copilots provide compelling, though nuanced, evidence regarding their influence on productivity and the distribution of performance. These field experiments offer insights into how generative AI could affect professional practice across sectors. Early empirical findings consistently show that access to generative AI tools can substantially increase output in diverse tasks. However, realizing these gains often depends on the user's capacity to engage effectively with the system. Productive interaction requires skill in prompt formulation, which involves clearly articulating task objectives, decomposing complex problems, and maintaining the metacognitive flexibility to revise strategies based on feedback \citep{Tankelevitch2024}. In laboratory settings, \citet{Li2024} demonstrate that improvements in user proficiency led to greater performance gains than upgrading from GPT-3.5 to GPT-4, indicating that human-AI collaboration skills are a key driver of value realization.

A multi-site trial involving over 6,000 users across sixty organizations found that integrating Copilot tools led to users completing documents 12\% faster and a reduction in time spent on email-related tasks \citep{Jaffe2024}. The study also documented spontaneous changes in organizational routines, such as delegating meeting-related tasks---including minute-taking and action-item tracking---to the AI system. However, these productivity gains were not uniformly distributed. Initial adoption patterns reveal substantial variation, even among firms actively pursuing generative AI integration. In a related study of more than 7,000 knowledge workers with access to an embedded AI assistant, researchers found that the average participant used the tool during 41 percent of study weeks. Firm-level averages ranged from as low as 6.3 percent to as high as 75 percent, highlighting significant heterogeneity in actual usage \citep{Dillon2025}. This variability suggests that factors beyond tool availability, such as job function, organizational support, and training, will likely influence adoption intensity and realized benefits.

In professional software development, three parallel randomized controlled trials involving 4,867 engineers demonstrated a 26 percent increase in merged pull requests when GitHub Copilot suggestions were used \citep{Cui2024}. This quantitative improvement was accompanied by a qualitative shift in work allocation, with engineering effort redirected from routine boilerplate coding toward higher-level design tasks. In a separate domain, a study conducted in a Fortune 500 call center found an average 15 percent increase in issue resolution per hour among agents using an AI copilot \citep{Brynjolfsson2025}. Similarly, in management consulting, professionals using AI completed 12.2 percent more tasks on average and did so 25.1 percent more quickly. Their outputs were also rated over 40 percent higher in quality than those of a control group without AI assistance \citep{DellAcqua2023Navigating}.

Further reinforcing these trends with newer AI modalities, a recent study by \citet{Schwarcz2025} on AI-assisted legal practice found that both an AI reasoning model (o1-preview) and a retrieval-augmented legal tool (Vincent AI) produced statistically significant improvements in the overall quality of legal work across most evaluated tasks. This represents a notable progression beyond earlier studies of models such as GPT-4, which reported more modest quality gains. In addition to these qualitative improvements, the study documented substantial productivity increases. Depending on the tool and the task, the AI systems raised quality points per minute by approximately 34 percent to 140 percent. These findings indicate that the frontier of AI-enabled support is shifting from basic efficiency gains to measurable improvements in output quality within complex professional contexts.

Moreover, early evidence suggests that the initial wave of productivity adjustments primarily occurs at the individual level. A large-scale study found that while AI usage led to substantial time savings on individual tasks, such as a 31 percent reduction in weekly email time for regular users, there were no statistically significant changes in time allocated to more collaborative activities, such as meetings over the six-month observation period \citep{Dillon2025}. These findings indicate that although individuals are optimizing their workflows in response to AI integration, broader gains in productivity that depend on shifts in coordination practices and organizational routines may emerge more gradually and require deliberate structural or managerial interventions.

Recent large-scale evidence from occupational deployments suggests that average productivity gains from generative AI may be relatively modest in the early stages. A comprehensive 2024 study conducted across eleven occupations in Denmark found that average user-reported time savings from AI chatbot use amounted to approximately 2.8 percent of total work hours. However, the observed effects varied considerably depending on the extent of employer encouragement and complementary organizational investments \citep{Humlum2025}. These findings reinforce the GAS framework’s implication that realizing the full benefits of generative AI requires more than tool adoption; it requires coordinated organizational change and sustained investment in complementary capabilities.

\subsection{Distributional Impact of Generative AI Adoption}

The uneven distributional effects of generative AI adoption illustrate how the GAS trade-off operates differently across skill levels. Novice users tend to benefit most from simplified access to broadly applicable capabilities, while expert users rely on domain-specific knowledge to manage the limitations in model accuracy.

A consistent pattern across multiple studies is that generative AI provides greater benefits to individuals with less experience or lower baseline performance, functioning as a skill-leveling tool. In a large-scale call center study, overall productivity increased by 15 percent, but novice agents experienced a substantially larger 36 percent improvement in resolution rates \citep{Brynjolfsson2025}. Similarly, a field experiment involving 758 consultants at Boston Consulting Group found that individuals below the average performance threshold improved by 43 percent when using GPT-4 for tasks within the model's capability frontier. In contrast, higher-performing peers benefited, though to a lesser extent, with a 17 percent performance increase \citep{DellAcqua2023Navigating}. In legal work, \citet{Schwarcz2025} reported that an advanced AI reasoning model (o1-preview) yielded greater improvements in output quality for law students with lower GPAs without diminishing performance among higher-achieving students. Moreover, a RAG-based legal AI tool exhibited more consistent quality improvements across skill levels, though productivity effects varied. 

These findings suggest that AI copilots can deliver immediate procedural scaffolding, stylistic modeling, and domain-specific assistance, thereby accelerating the upskilling of less experienced users. For junior developers, this may mean access to idiomatic code patterns; for early-stage entrepreneurs, the rapid generation of foundational documents. However, the same studies underscore that novice users remain more susceptible to over-reliance on AI outputs and may struggle with metacognitive calibration, increasing the importance of targeted training and oversight \citep{Tankelevitch2024}. This over-reliance can be hazardous in areas like software development, where novices may lack the experience to spot plausible-but-insecure code. As a comparative study by \citet{hamer2024another} noted, both AI-generated code and human-written code from forums like StackOverflow contain vulnerabilities, but the types of errors can differ, requiring nuanced expertise to detect.

While novices often realize the largest relative gains, the impact of generative AI on high-skilled or top-performing individuals is also substantial, typically manifesting as increased output or the capacity to address more complex challenges. In the consultant field experiment, higher-performing individuals improved their performance scores by 17 percent when using GPT-4, despite already strong baseline capabilities \citep{DellAcqua2023Navigating}. Similarly, research on entrepreneurial founders showed that those in the top quartile of ex-ante ability translated AI assistance into a 15 percent increase in revenue \citep{Otis2024}. However, the same study revealed that founders in the bottom quartile experienced a decline in performance when using AI, likely due to challenges in detecting and overriding inaccurate or misleading outputs. These findings indicate that while AI tools can augment the capabilities of top performers, the benefits depend on users' ability to evaluate and integrate AI-generated content critically. As tasks approach the limits of model accuracy, human judgment remains essential---a direct reflection of the constraints imposed by the accuracy ceiling within the GAS framework.

\subsubsection{Variation by Task Type and Domain Complexity}

\citet{RoldanMones2024} examined the impact of generative AI support in a university debating competition, a task characterized by unpredictable interactions and a reliance on higher-order reasoning skills. The study found that higher-ability students derived substantially greater benefits from ChatGPT assistance. These students were more effective at extracting and applying AI-generated content, particularly in dimensions such as argument credibility and refutation. In contrast, lower-ability participants experienced modest gains, primarily in surface-level aspects such as presentation clarity. The authors argue that when tasks require substantial human judgment to navigate the limitations imposed by the AI's accuracy ceiling, especially in dynamic, non-formulaic settings, pre-existing cognitive and rhetorical skills complement effective AI use.

Evidence from educational interventions in developing countries further illustrates the uneven distributional effects of generative AI. \citet{DeSimoneEtAl2025} evaluated a virtual English tutoring program for secondary students in Nigeria that utilized GPT-4. While the intervention produced significant overall learning gains, the most considerable improvements were concentrated among students with higher initial academic performance. These results indicate that realized benefits from AI-enhanced learning may be mediated by prior academic preparation, digital literacy, access to supportive infrastructure, and a greater capacity to engage effectively with AI-based educational tools.

The underlying reasons for enhanced performance among experienced individuals, particularly in technical domains such as software development, often stem from their active and discerning engagement with AI-generated content rather than passive acceptance. \citet{Osmani2024} notes that experienced engineers use AI most effectively by applying their expertise to guide, refine, and constrain the model's output. This includes iterative practices such as refactoring generated code into modular components, addressing overlooked edge cases, reinforcing type definitions, scrutinizing architectural choices, and implementing robust error handling. In this context, AI accelerates tasks the user understands well. It enables rapid prototyping and automation of routine components while the engineer’s domain knowledge ensures the final product is accurate, reliable, and maintainable. In contrast, less experienced users may be more inclined to accept suggestions at face value, increasing the risk of deploying flawed code. Ultimately, the accuracy of AI-assisted work is not solely a function of model capability but also of the human expert’s capacity to critically assess and refine the output.

These field experiments provide empirical support for the Generality–Accuracy–Simplicity (GAS) framework. Generative AI systems lower the \emph{Simplicity} threshold by embedding advanced capabilities within accessible, conversational user interfaces. They simultaneously expand the \emph{Generality} domain, enabling a single system to address heterogeneous tasks, including text generation, code synthesis, summarization, and basic reasoning. However, the persistence of the \emph{Accuracy} constraint remains evident. In domains where the tolerance for error is high, such as drafting preliminary responses, generating boilerplate code, or assisting with routine consulting tasks, less experienced users can realize substantial productivity gains.

In contrast, tasks characterized by high-stakes decision-making or production-critical outputs, such as strategic entrepreneurship or software development at scale, continue to require expert human oversight to mitigate risks arising from LLM limitations. As a result, generative AI functions as a capability amplifier, with utility often inversely correlated with task criticality. In response, organizations are beginning to restructure workflows and talent deployment: junior personnel are increasingly supported by AI tools to boost output, mid-career professionals are being retrained to emphasize judgment and quality assurance, and senior experts are tasked with designing and governing AI-integrated systems that manage both technical and organizational complexity.

\subsection{Economic Implications of Reduced Cognitive Cost}
\label{economic-implications-llms-as-a-price-shock-in-general-purpose-cognition}

The introduction of large-scale language models (LLMs) represents an important shift in the economics of cognitive work. By combining broad \emph{Generality} with user-facing \emph{Simplicity}, these systems substantially reduce the marginal cost of producing functional outputs such as text, code, and summaries. Tasks that previously required coordinated effort across multiple roles can often be completed with a single prompt at minimal cost. This reduction enables new service delivery approaches, particularly in areas where traditional labor-intensive methods are inaccessible. For example, Garfield AI, a UK-based firm approved by regulators, now offers legal services such as debt collection at approximately \pounds 2. This case illustrates how generative AI may lower barriers in cost-sensitive domains, potentially broadening access to previously difficult services to scale economically \citep{Ring2025GarfieldAI}.

This potential is observable at both individual and organizational levels. For example, Pete Koomen's conceptual “inbox experiment” illustrates how a conversational AI, when guided by a personalized system prompt, might streamline email triage, summarization, and draft preparation into a single, cohesive workflow, offering more than just isolated draft generation and edging toward integrated email management \citep{Koomen2025}. Similarly, Matthew Sinclair reports generating approximately 30,000 lines of code over a few weeks for both frontend and backend projects using Claude Code, a volume of work he estimates would have taken months using traditional methods \citep{Sinclair2025}. These examples suggest a significant decline in the marginal cost of diverse cognitive tasks. As such, it may be appropriate to view large language models as contributing to a marked reduction in the cost of general-purpose knowledge work.

\subsubsection{Standard Economic Responses to Input Cost Decline}
\label{standard-responses-to-a-cost-decline}

Classical price theory predicts four effects when the unit cost of an input declines: (a) quantity demanded increases, (b) new applications emerge, (c) substitutes may lose value, and (d) complements may become more valuable. Early deployment of large language models (LLMs) shows each of these dynamics in action:

\begin{itemize}
\item
\textbf{Higher volume:} The low marginal cost of producing content with general-purpose LLMs, especially in domains where high Accuracy is not essential, has led to a substantial rise in output volume. It is now common for teams to generate multiple headline variants, large batches of code suggestions, or expansive sets of test cases overnight, enabled by the negligible cost per draft.

\item
\textbf{New applications:} Lower costs also make previously impractical use cases feasible, as AI significantly reduces the search cost to discover novel, high-value combinations \citep{AMO2018}. Klarna’s large-scale content personalization, AI-driven onboarding tutorials, automated compliance briefings, and synthetic user stories are examples of applications that were unlikely to be cost-effective when each draft required manual effort but are now viable at current token pricing.

\item
\textbf{Displacement of substitutes:} As general-purpose models become more capable, specific point solutions and specialist roles face competitive pressure, especially those that do not require high Accuracy. For example, generative AI has reduced demand for freelance tasks in online labor markets, prompting shifts in how workers position their services \citep{Yiu2025}. This trend is more pronounced in structured, repetitive environments (what \citet{Epstein2019} terms “kind” learning environments). In contrast, human expertise often retains value in less predictable, “wicked” domains and may even become more essential.

\item
\textbf{Valuation of complements:} As the ability to generate content becomes widespread, the ability to critically evaluate, refine, and contextualize that content gains importance. Roles involving quality assurance, risk assessment, and strategic interpretation may become bottlenecks in LLM-enabled workflows. These tasks often require strong metacognitive skills and the capacity to assess one’s judgment and adjust confidence appropriately, increasing the value of human oversight \citep{Tankelevitch2024}. Consistent with this, an analysis of job postings found that the wage premium for possessing AI skills was highest not for pure technologists, but for managers, indicating that the ability to combine AI with organizational strategy and business judgment is exceptionally valuable to employers \citep{Alekseeva2021}. In high-stakes or ambiguous settings, such as those described by \citet{Epstein2019}, individuals with a broad range of experiences and interpretive frameworks are likely to be especially well-positioned to add value.
\end{itemize}

The growing prevalence of AI-generated errors, often delivered with unwarranted confidence due to issues like the ``flattening calibration curve,'' where models appear more certain even when incorrect \citep{VarianceLabs2025}, intensifies the need for human oversight. Businesses are increasingly focused on managing and reducing the risks associated with such inaccuracies, which elevates the importance of human expertise in verification, critical assessment, and nuanced contextual judgment.

Creating new tasks is a critical way for automation technologies to sustain or even increase labor demand. Supporting this, \citet{Humlum2025} found that AI chatbots led to new job tasks for 8.4 percent of workers, including some who did not use the tools directly. These new responsibilities often included integrating AI into workflows, reviewing AI outputs, and addressing ethics and compliance. Such activities complement the deployment of general-purpose AI. 

This shift also contributes to a structural change in the AI industry. Building and maintaining leading foundation models now requires substantial resources and capabilities. As a result, some firms specialize in creating and distributing these general-purpose systems, while others focus on developing tailored applications in various parts of the `stack.' This division allows organizations to leverage foundation models without needing in-house research and development, accelerating the spread of AI-enabled solutions throughout the economy.

These economic responses are amplified by a powerful behavioral dynamic. Building on Tesler's Law, Bruce Tognazzini (\citeyear{Tognazzini1998}) noted that as a given task is simplified, users do not cease their efforts; instead, they tend to take on more challenging and complex tasks, effectively maintaining a consistent level of cognitive engagement. This `Complexity Paradox' suggests that the reduction in the cost of generalist cognition will not lead to a static endpoint, but rather to a continuous expansion of the G-A frontier. As users become accustomed to current AI capabilities, they will demand tools to solve ever-more-complex problems, ensuring that the burden of hidden complexity absorbed by organizations and their technical teams will only increase over time.

\subsection{Labor Implications: Generalists, Specialists, and the Automation Frontier}

\subsubsection{Generalist versus specialist human capital.}
\label{generalist-versus-specialist-human-capital}

\citetposs{Dhar2016} Decision-Automation Map plots achievable accuracy against the cost of error, defining an ``automation frontier'' above which purely machine decisions remain unacceptable. The LLM price shock shifts many tasks rightward toward greater automation, but only up to that frontier. 

Formal economic models provide a rigorous foundation for these labor market shifts. Using a model of knowledge hierarchies, \citet{Ide2025} demonstrate that AI's impact depends critically on its autonomy and knowledge level. Their model predicts that non-autonomous "co-pilots" tend to benefit less knowledgeable workers the most, whereas autonomous "co-workers" primarily benefit the most knowledgeable individuals, providing a theoretical mechanism for the diverging value of different skills.

\begin{itemize}
\item
\textbf{Generalists} who can orchestrate LLM output across heterogeneous domains, integrate partial answers, and frame the next query increase their productivity disproportionately. This aligns with the value of ``range,'' where diverse experiences allow individuals to make novel connections and adapt to the challenges posed by AI integration. Research has shown that individuals with greater knowledge diversification (generalists) are more adept at integrating new, distant knowledge external to their existing expertise, effectively becoming ``masters of knowledge'' in novel domains \citep{Nagle2020}. Technology directly complements their labor, particularly in navigating complex, ill-defined problems where narrow specialists may struggle.
\item
\textbf{Specialists} retain their wage premium and, in high-risk sectors, may even see it extended, but only where accuracy requirements remain above the automation frontier (such as in safety-critical engineering, clinical medicine, or actuarial modeling). In lower-risk areas of narrow routine expertise, that premium tends to erode, a trend supported by evidence showing decreased demand for traditional managerial skills like budgeting and scheduling in firms with high AI adoption \citep{Alekseeva2024}. Research on elite career paths has found that specialization is consistently rewarded, but for different reasons at different career stages: early on, it serves as a signal of general ability, while later it reflects the substantive value of the specific skills accumulated \citep{Ferguson2013}.
\end{itemize}

Optimal human capital, therefore, tends toward a T-shape: broad, prompt-literate capability and the ability to draw from diverse knowledge domains (range), combined with a deep spike of domain knowledge that protects the accuracy boundary. This capacity to explore and integrate external knowledge suggests that individuals with range are well positioned to develop specialized depth when needed \citep{Nagle2020}. Such breadth is often cultivated through a sampling period of varied experiences, which is becoming increasingly important as AI systems automate more specialized and routine tasks.

This T-shaped model is consistent with observations from freelance labor markets. Higher-skilled specialists appear less likely to make costly horizontal moves into entirely new domains following the adoption of AI tools. Instead, they often adapt by repositioning vertically, such as targeting different types of job value or by adding AI-related capabilities to their profiles. These strategies allow them to preserve the relevance of their expertise while adjusting to new competitive dynamics \citep{Yiu2025}.

The relative advantage of generalists and specialists also depends on the pace of change within a given knowledge domain. In slower-moving fields, generalists perform well by making connections across areas and introducing novel combinations of ideas. In rapidly evolving domains, however, specialists can have an edge because their expertise allows them to absorb and apply new knowledge more quickly when it emerges within their area \citep{Teodoridis2019}. As AI accelerates change in some domains and flattens access to foundational knowledge in others, understanding this dynamic becomes essential for workforce strategy.

The shift toward AI-driven efficiency also introduces a challenge to traditional models of expertise development. If early-career tasks that once enabled novices to learn from experts are automated or devalued, the pathways to becoming either a generalist or a specialist may become more limited. This could lead to long-term talent shortages and higher costs for cultivating the advanced conceptual skills that organizations will continue to need.

\subsubsection{Conceptual Reasoning as a Scarce Complement}
\label{conceptual-reasoning-versus-practical-experience}

Because large language models (LLMs) are statistical pattern completers rather than structured reasoners, their most serious errors often occur when outputs appear plausible but violate fundamental principles. Identifying such issues depends on conceptual reasoning skills, such as abstraction, causal thinking, and identifying counterexamples. As a result:

\begin{itemize}
\item
\textbf{Conceptual skills}, including problem framing, formal modeling, and edge-case identification, are becoming more valuable. These skills help organizations remain within acceptable risk boundaries when using AI tools. Applying them effectively in an AI-augmented environment also requires an understanding of when and how to incorporate AI. This involves a higher-level metacognitive capacity, such as assessing when AI is appropriate and how to manage its output \citep{Tankelevitch2024}. While AI can accelerate routine work, developing deeper conceptual insight often requires engaging with challenging problems. This learning process may be slower, but it is important for long-term adaptability. In addition, as AI affects more individual workflows than team-based practices initially \citep{Dillon2025}, an emerging area of value is the ability to adjust collaborative processes in light of new technologies.
\item
\textbf{Routine experience} that once signaled competence, such as writing boilerplate code or formatting reports, may now be partially replaced. Entry-level employees are increasingly expected to refine or review AI-generated content rather than produce it from scratch.
\item
\textbf{Career development paths may be changing}: early roles are more likely to involve working with AI outputs. In contrast, progression involves stronger conceptual judgment and the ability to guide and improve automated contributions.
\end{itemize}

Overall, the declining cost of general-purpose AI capabilities may lead to anticipated shifts in work patterns. In particular, we should expect to see increases in volume, more use cases, changes in demand for specific skills, and a growing demand for human abilities that complement AI systems. The relative value of human work is increasingly found in areas that require high accuracy, domain knowledge, or abstract reasoning, areas that remain just beyond what current AI systems can reliably handle.

\section{Rethinking Expertise and Capability Development}
\label{sustaining-the-expertise-pipeline-in-an-llm-mediated-workplace}

The separation of complexity through abstraction, discussed in Section 2.1, presents a challenge for developing expertise. This challenge is a consequence in part of the Law of Leaky Abstractions; because the simplified interfaces of AI will inevitably fail or `leak,' practitioners who do not understand the underlying principles will be unable to diagnose or solve critical problems \citep{Spolsky2002Leaky}. As large language models (LLMs) increasingly conceal complexity behind simplified interfaces, organizations must intentionally design pathways for practitioners to build the conceptual understanding needed to manage that hidden complexity.

One unintended consequence of the sharp drop in the cost of general-purpose cognition is that it may disrupt traditional routes to deep expertise. In the legal field, for instance, leaders worry that the automation of routine tasks like basic research, cite-checking, and first drafts that were ``once rites of passage for junior attorneys'' will erode the foundational training ground for the next generation. This challenge makes the creation of new, deliberate practice frameworks a leadership imperative \citep{thomsonreuters_legal_2025}.

As \citet{Osmani2024} notes, generative AI tools can benefit experienced users more than novices, raising the risk that early-career professionals will plateau before acquiring the judgment needed to assess machine-generated output critically. This concern is reinforced by research in cognitive psychology showing that skills decay more quickly than expected once they are no longer practiced. A meta-analysis of 189 effect sizes found substantial performance declines after just one year of nonuse \citep{Arthur1998}.

This dynamic has sparked concern among technical professionals that the industry risks ``robbing ourselves of good future developers'', in the words of a Hacker News discussant\footnote{https://news.ycombinator.com/item?id=44167785}, as the very tasks that build foundational expertise are automated away. The anxiety is sharpened by the provocative framing from some AI proponents that a ``\$20/month'' subscription to a coding agent can be viewed as a replacement for a human intern \citep{Ptacek2025Nuts}, raising existential questions about the future viability of entry-level roles.

Emerging evidence in educational settings suggests that while generative AI can enhance performance in the short term, it may also hinder underlying skill development if not used thoughtfully. In one study, \citet{Bastani2024} found that students who used a standard GPT-4 interface to solve math problems performed worse on follow-up exams without AI than peers who had not used AI. However, these adverse effects were largely mitigated when students interacted with a modified version of GPT-4 designed to support learning by offering incremental hints instead of direct answers.

\subsection{Lessons from Automation in High-Reliability Domains}

The aviation sector offers a relevant parallel. Before the rise of generative AI, the industry recognized the risks of over-reliance on automation. Pilots’ hands-on flying and troubleshooting abilities diminished without regular practice. As a response, flight simulators became a critical tool for sustaining manual competence. As \citet{Sellen2024} argue, the ``AI co-pilot'' metaphor should be accompanied by design choices that keep users engaged, maintain their decision-making authority, and support the development of human skills. Without these measures, the convenience of automation may come at the expense of long-term capability.

A compounding factor in AI adoption is that effective use of these tools requires developing advanced metacognitive skills, such as self-awareness, confidence calibration, and task decomposition, which must be intentionally cultivated \citep{Tankelevitch2024}. Research highlights that AI use can lead to overconfidence or a mismatch between perceived and actual performance. For example, \citet{Fernandes2024} found that users with higher technical AI knowledge sometimes showed less accurate self-assessments despite greater confidence. These findings suggest that developing fluency with AI tools should go hand in hand with training in self-monitoring and evaluation.

This challenge is further illustrated in ``shadow learning,'' where individuals use AI to complete tasks but miss opportunities to practice and develop underlying skills. In programming education, \citet{Prather2024} found that AI tools may widen performance gaps by helping already capable students advance while inadvertently reinforcing misconceptions among struggling students. The simplicity of getting answers from AI can mask a lack of understanding if students do not actively engage with or question the output.

In software engineering, similar dynamics are emerging. \citet{Osmani2024} observes that while experienced developers benefit from AI by steering and refining its suggestions, junior developers risk missing out on learning key skills such as debugging, architectural thinking, and error handling. The so-called ``70\% problem'' arises when AI tools help complete the bulk of a task but leave the most difficult and instructional aspects unfinished. This makes it harder for less experienced users to develop the depth of understanding needed to solve complex problems or maintain high-quality systems independently.

To address this, some experts advocate focusing on ``durable skills'' that are less likely to be displaced by AI \citep{Osmani2025}. These include tasks that require system-level reasoning, critical evaluation of ambiguity, and long-term contextual awareness. Such skills are not easily automated and are essential for ensuring quality and reliability in complex work.

Practitioners are already raising concerns. Blogger \citet{Sinclair2025} notes that developers without much hands-on experience may fail to notice flawed AI outputs, particularly when they lack the deeper architectural knowledge to evaluate results properly. Engineers on forums like Hacker News report that junior colleagues raised on AI tools sometimes struggle to build or debug basic structures independently.\footnote{https://news.ycombinator.com/item?id=43752492} In a controlled experiment, humans collaborating with AI agents decreased their direct text editing by 60\% but increased their communication with the AI by 137\%, focusing more on high-level suggestions, instructions, and planning \citep{Ju2025Collaborating}, suggesting a reallocation of human effort away from hands-on execution, and the deliberate practice it entails, toward a more managerial and directive role. Without careful attention to training and skill development, the pool of professionals capable of handling high-accuracy work could shrink just as demand for this expertise is rising.

Generalist skills may help mitigate this challenge. Research suggests that individuals with broad experience across domains are better equipped to integrate unfamiliar information, adapt to new challenges, and develop expertise in emerging areas \citep{Epstein2019, Nagle2020}. While specialists retain an advantage in rapidly evolving fields, generalists can provide valuable flexibility, particularly when tasks or domains shift. Encouraging range in experience may help ensure that teams remain adaptable and can cultivate new expertise as the AI landscape evolves \citep{Teodoridis2019}.

Professional training programs are also beginning to respond to these dynamics. In legal education, there is ongoing discussion about how AI support might affect the development of critical thinking and analytical reasoning. If students rely too heavily on AI tools to draft or analyze arguments, they may not build the skills required for independent legal reasoning unless educational structures are adjusted \citep{Schwarcz2025}.

Organizations are responding in kind. For example, law firm Crowell \& Moring has begun introducing generative AI tools selectively, starting with adjacent legal tasks and extending to core work in situations where risks can be managed. They also introduced required AI training for staff, recognizing the need to build proficiency and judgment as AI tools become more common \citep{Hammond2024USLawFirms}.

Evidence from large-scale studies reinforces the importance of structured implementation. \citet{Humlum2025} found that employees reported stronger gains in productivity, quality, and creativity when AI adoption was paired with active support from employers, including training programs and in-house tools. These results suggest that thoughtful deployment, aligned with the GAS framework, depends on technical integration and investments in human capability.

\subsection{Design Principles for Maintaining Deliberate Practice}
\label{design-principles-for-preserving-deliberate-practice}

The following design principles aim to counteract the erosion of deliberate practice and support the development of robust expertise pipelines in an LLM-mediated workplace. Their importance is underscored by findings that the efficacy of generative AI is not solely a function of technological sophistication. For instance, laboratory evidence suggests that upgrading from GPT-3.5 to GPT-4 yields only modest performance gains unless users receive targeted collaboration training, proving more impactful than the model upgrade \citep{Li2024}.

\begin{enumerate}
\def\labelenumi{\arabic{enumi}.}
\item
\textbf{Developing Prompt Engineering Skills.} Training individuals to formulate clear and effective prompts is essential. As LLMs become more deeply embedded in workflows, the ability to articulate problems and desired outputs becomes a key competency. This skill boosts productivity and helps users generate more accurate and reliable results from models \citep{Denny2024}. It promotes better engagement with systems that may appear simple but require carefully crafted input to perform well.

\item
\textbf{Progressive Disclosure of Automation.} Organizations should consider gating LLM features behind demonstrated competency milestones. For example, new hires might manually complete small assignments or documentation and gain access to AI tools only after demonstrating foundational proficiency. This staged exposure helps users develop first-principle reasoning skills before relying on AI assistance. It also supports learning through retrieval practice and self-explanation, essential to building long-term understanding \citep{Dunlosky2013}.

\item
\textbf{Dual-Track Deliverables.} Each AI-assisted output should be accompanied by a short human-written rationale that explains why the result is plausible and outlines conditions where it might fail. This requirement promotes reflective practice and encourages users to critically evaluate the AI’s suggestions, enhancing metacognitive awareness and judgment \citep{Tankelevitch2024}.

\item
\textbf{Rotation Through Low-Automation Environments.} Regular assignments in settings where LLM use is restricted, such as incident response, security-critical codebases, or formal academic exercises, help preserve essential skills. These experiences maintain familiarity with fundamentals that automated tools cannot reliably support.

\item
\textbf{LLM as Tutor, Not Oracle.} AI tools intended for training should return Socratic prompts, such as ``What alternative hypothesis could invalidate this conclusion?'' rather than final answers. This helps scaffold inquiry and promotes critical thinking. Research supports this approach: reasoning models can provide specific, rubric-guided feedback on complex tasks such as legal writing or technical explanations, offering scalable and individualized support \citep{Schwarcz2025}. Additional studies show that AI tools designed with pedagogical safeguards, like offering incremental hints instead of direct answers, can mitigate adverse impacts on learning \citep{Bastani2024}. \citet{Lehmann2024} found that students benefited from using LLMs as personal tutors for explanations, while overreliance on AI for task completion without active engagement impaired learning outcomes. \citet{Vanzo2024} found that an AI tutor that provided interactive, step-by-step guidance without revealing direct answers led to significant improvements in student learning outcomes and engagement, especially for those with weaker initial skills.

\item
\textbf{Structured Review Apprenticeships.} Junior staff should work alongside experienced reviewers when evaluating AI-generated content. This pairing helps them learn to spot subtle issues such as hallucinations, reward optimization artifacts, or silent logic violations. Professional heuristics cannot be easily codified and are best transmitted through mentorship. For such collaborations to succeed, foundational training is essential, this includes understanding the strengths and limitations of various AI systems (e.g., reasoning models versus RAG-enabled tools) and being aware of ethical considerations relevant to their use in professional environments \citep{Schwarcz2025}.
\end{enumerate}

The nature of the AI's interaction style may also influence skill development. While current LLMs are typically designed for supportive and straightforward interactions, there is growing interest in the idea of ``antagonistic AI.'' This concept involves models that deliberately challenge users, which may help foster critical thinking and reduce over-reliance. These systems could support the development of evaluative expertise, but they would require careful ethical considerations, including explicit user consent and thoughtful implementation \citep{Cai2024}.

Recent research shows that the `personality' of the AI, as determined by its system prompt, interacts with the human user's personality to significantly affect performance. For example, pairing a conscientious human with an AI agent prompted to be `open' improved creative output, whereas pairing an extroverted human with a `conscientious' AI degraded quality \citep{Ju2025Collaborating}. This suggests that a crucial, and perhaps ultimate, design principle involves tailoring the AI's collaborative style to complement the human user, creating a more effective and synergistic partnership.

In software development, \citet{Osmani2024} proposes a hybrid strategy for users of AI coding tools, especially non-engineers. The approach includes using AI for rapid prototyping, taking time to understand the generated code, learning basic programming concepts alongside AI use, gradually building a foundation of knowledge, and using AI as a learning aid rather than just a code generator. This method highlights the importance of engaging actively with AI-generated content to ensure meaningful learning and skill development.

Universities and professional schools face similar challenges. Curricula that previously ended with automated capstone projects may now need to begin with automation-aware training. This includes teaching prompt design, understanding how errors propagate, and analyzing the potential consequences of mistakes. Manual derivation exercises should be followed to reinforce the depth of understanding. Assessments should prioritize conceptual thinking and abstraction, such as reasoning through algorithmic trade-offs or causal relationships, instead of focusing on routine output that AI tools can produce. The goal is to prepare students to question and verify AI outputs before relying on them in professional practice.

\citet{Kasneci2023} emphasize that although LLMs can be valuable in education, they also require new competencies for students and educators. A significant concern is the risk that learners may become overly dependent on AI, which can hinder the development of independent problem-solving skills. Educational strategies must focus on using AI to complement, not replace, the learning process.

LLMs have significantly reduced the cost of producing general cognitive work. However, this gain may be short-lived if organizations neglect the development of human expertise. To sustain long-term value, workplaces, and educational institutions must incorporate strategies that promote progressive independence, require thoughtful justification, and ensure that individuals regularly practice core skills without relying solely on automation. These efforts help preserve the human judgment and accuracy that remain essential in many high-stakes applications.

\subsection{Evolving Frontiers: Complexity, Capability, and Substitution}
\label{the-moving-frontier-how-rising-model-complexity-reshapes-complements-and-substitutes}

The GAS trade-off is not static. Each training run that uses more parameters, longer context, or richer multimodal data increases hidden complexity and, as a result, pushes the generality–accuracy frontier outward. Over time, tasks that once stood safely above the automation boundary move into the zone where machine assistance becomes viable. The balance between human roles that complement AI and those replaced by it changes accordingly.

This frontier is advancing toward greater capability in complex tasks. Research by \citet{Schwarcz2025} on the impact of newer AI systems, specifically reasoning models such as o1-preview and retrieval-augmented generation (RAG) tools like Vincent AI, shows that these technologies can improve both efficiency and the quality and analytical depth of legal work produced by students. This marks a meaningful shift because earlier models primarily delivered speed improvements. Although the reasoning model continued to produce hallucinations, the RAG system was more successful in limiting them. This suggests that with targeted system design, extending the accuracy frontier in specific domains is possible, even if general-purpose models continue to face reliability challenges. These findings show how new tools are beginning to automate or significantly enhance tasks that previously depended on intensive human legal judgment.

However, the movement of the frontier does not always proceed in a straight line or deliver unambiguous gains. Some recent advanced reasoning models have shown increased rates of hallucination compared to their predecessors.\footnote{https://www.nytimes.com/2025/05/05/technology/ai-hallucinations-chatgpt-google.html} This illustrates that improvements in one area can introduce setbacks in another. Although the potential for automation continues to expand, the shape of the accuracy ceiling may shift in unexpected ways, revealing new types of error or making older issues more challenging to manage.

\subsubsection{Machine Translation as a Case of Frontier Advancement}
\label{a-case-in-point-translation}

Machine translation has seen remarkable advancements. While earlier neural machine translation systems delivered fluent renditions of newswire prose, they often faltered on the complexities of legal depositions, medical reports, or ancient texts like Homeric Greek. These tasks required contributions from professional domain experts for post-editing jargon, preserving meter, and correcting for cultural nuance. As model complexity has increased, translations' generality and accuracy have improved considerably. For instance, GPT-4 has demonstrated performance comparable to junior-level human translators' overall error rates across various domains and maintains consistent translation quality even in resource-poor language directions where traditional translation systems struggle \citep{Yan2024}.

Recent studies indicate that such advanced LLMs are ``decisively better and more closely approaching human literary translation'' than previous systems. In literary translation, systems like GPT-4o rank closely behind professional human translators. While human translations are still generally preferred for their nuanced understanding of literary style and cultural context, the capabilities of LLMs have advanced to a point where they can produce high-quality translations of complex literary works \citep{Zhang2025}. This narrows the set of tasks exclusively reliant on human experts. However, nuanced areas like capturing aesthetic value, avoiding overly literal renderings and ensuring lexical diversity still benefit from human oversight. Indeed, even experts can find it challenging to consistently distinguish between high-quality LLM and human translations, indicating that the frontier of AI capability in this domain continues to expand rapidly \citep{Yan2024, Zhang2025}.

The pattern generalizes. Any fixed task tends to follow a predictable arc:
\begin{itemize}
\item
\textbf{Early stage:} Accuracy is low; domain expertise is essential.
\item
\textbf{Middle stage:} Rising complexity improves accuracy; demand concentrates on a smaller group of higher-level specialists who correct subtle edge cases.
\item
\textbf{Late stage:} If complexity continues to scale, machine accuracy approaches specialist performance, and even rare expertise becomes substitutable.
\end{itemize}

In the limit of unbounded scale, toward an AGI horizon, specialist human abilities could become fully substitutable.

\textbf{Judgment remains a complement.} Even as model-based complexity improves generality and accuracy, assigning translations (or any output) to real-world purposes still requires human decision-making. This includes selecting which rendering best captures a client's voice, deciding when legal nuance requires literal fidelity, or determining whether a meter-perfect Iliad is more pedagogically useful than a freer rendition. These decisions depend on human judgment, which relies on metacognitive processes such as confidence calibration and strategy selection \citep{Tankelevitch2024}. Evaluating sufficiency, alignment with intent, and accepting residual risk remains a human responsibility. Accordingly, while complexity steadily converts many domain-specific skills from complements into substitutes, deciding what ``good enough'' means in context remains a high-value human contribution at the frontier.

As the generality-accuracy frontier shifts and AI takes on more specialized, predictable tasks, the premium on human skills will likely concentrate among those who can navigate ambiguity, integrate AI outputs with context, and exercise judgment where AI falters. This shift emphasizes the rising value of broad expertise and conceptual reasoning.

\section{Dynamic Consequences for Inequality and Evaluation}
\label{dynamic-implications-for-inequality-and-evaluation-studies}

Because tasks move along the frontier over time, the distributional effects of LLM adoption evolve:

\begin{itemize}
\item
\textbf{Short run:} Inequality tends to rise. Early adopters with scarce complementary skills (e.g., prompt engineering, high-level review) capture disproportionate value.
\item
\textbf{Medium run:} As complexity improves automation, these rents erode, and mid-tier roles lose some of their premium.
\item
\textbf{Long run:} Once accuracy saturates, fewer human contributions are needed, and task-specific inequality declines.
\end{itemize}

This means that today's empirical studies of productivity gains (such as with customer-support copilots or junior research assistants) capture only a snapshot of the technology's maturity. Each technical improvement repositions tasks along the GAS surface, turning some human roles from complements into substitutes and prompting new evaluations of winners and losers.

Furthermore, these changes often come from model updates that produce unexpected shifts in behavior. A new version may improve performance and introduce a distributional shift that undermines previous assumptions or invalidates existing system designs \citep{VarianceLabs2025}. This means that workflows, human-AI complementarities, and even established safety assessments may require rapid reevaluation. Organizations must be prepared to adjust their deployment strategies and oversight practices to keep pace with these often unpredictable developments.

The result is a reconfiguration of work. Generative AI increases the volume and diversity of output, making judgment under uncertainty the most scarce and valuable human skill. Successful organizations will map workflows to their appropriate place along the frontier, implement safeguards where mistakes are costly, and use the surplus of AI-generated drafts to support carefully filtered human reviews. For individuals, career leverage now lies upstream. The ability to frame problems, identify exceptions, and intervene when a model veers off course is becoming more critical than raw production. LLMs do not eliminate work. Instead, they reorganize it around the expanding generality-accuracy frontier.

\section{Conclusions}

Large language models appear to offer a free lunch --- broad generality and high accuracy behind an intuitive interface --- but the GAS framework reveals this to be an illusion. The simplicity experienced by the user is sustained only because immense technical and organizational complexity has been absorbed by hidden layers of infrastructure, process, and talent; the fundamental trade-off is not abolished, but merely relocated. This relocated complexity manifests most visibly as a persistent ``accuracy ceiling'', where residual errors become a fundamental constraint, forcing organizations to decide how much complexity they can shoulder and where to re-insert human judgment to ensure reliability.

Strategic advantage, therefore, shifts from a narrow focus on marginal model improvements to the architectural challenge of designing workflows that operate at the correct point on the Generality-Accuracy frontier. The practical implications are clear: tasks must be mapped according to their tolerance for cost, latency, and error. This means reserving lightweight models for routine work, while deploying more complex, retrieval-augmented systems for high-stakes functions, and embedding human-in-the-loop verification where the cost of failure is high.

Ultimately, this turns the GAS constraint from a limitation into a core design principle. It provides a blueprint for aligning technology, governance, and talent development with the reality that scalable cognition is most valuable when an organization masters the complexity it hides. The key to success is not just adopting AI, but managing the trade-offs it imposes.

\newpage
\singlespacing
\bibliographystyle{plainnat}
\bibliography{gas}

\end{document}